\documentclass[%
 reprint,
superscriptaddress,
 amsmath,amssymb,
 aps,
 prb,
]{revtex4-2}

\usepackage{xcolor}
\usepackage{graphicx}%
\usepackage{dcolumn}%
\usepackage{bm}%
\usepackage{hyperref}%
\usepackage{multirow}

\begin{document}

\preprint{APS/123-QED}

\title{Lessons from the harmonic oscillator -- a reconciliation of the Frequency-Resolved Frozen Phonon Multislice Method with other theoretical approaches}%

\author{Paul M. Zeiger}
\email{paul.zeiger@physics.uu.se}
\affiliation{%
  Department of Physics and Astronomy, Uppsala University, P.O. Box 516, 75120 Uppsala, Sweden
}%

\author{Juri Barthel}
\affiliation{Ernst Ruska-Centre (ER-C 2) Forschungszentrum Juelich GmbH, 52425 Juelich, Germany}

\author{Leslie J. Allen}
\affiliation{School of Physics, University of Melbourne, Parkville, VIC 3010, Australia}

\author{J\'{a}n Rusz}%
\affiliation{%
  Department of Physics and Astronomy, Uppsala University, P.O. Box 516, 75120 Uppsala, Sweden
}%

\date{\today}%

\begin{abstract}
We compare the Frequency-Resolved Frozen Phonon Multislice (FRFPMS) method, introduced in Phys. Rev. Lett. \textbf{124}, 025501 (2020), with other theoretical approaches used to account for the inelastic scattering of high energy electrons, namely the first-order Born approximation and the quantum excitation of phonons model. We show, that these theories lead to similar expressions for the single inelastically scattered intensity as a function of momentum transfer for an anisotropic quantum harmonic oscillator in a weak phase object approximation of the scattered waves, except for a too small smearing of the scattering potential by the \emph{effective} Debye-Waller factor (DWF) in the FRFPMS method. We propose that this issue can be fixed by including an \emph{explicit} DWF smearing into the potential and demonstrate numerically, that in any realistic situation, a FRFPMS approach \emph{revised} in this way, correctly accounts for the single inelastically scattered intensity and the correct elastic scattering intensity. Furthermore our simulations illustrate that the only requirement for such a \emph{revised} FRFPMS method is the smallness of mean squared displacements for all atomic species in all frequency bins. The analytical considerations for the FRFPMS method also explain the $1/\omega^2$-scaling of FRFPMS spectra observed in Phys. Rev. B \textbf{104}, 104301 (2021) by the use of classical statistics in the molecular dynamics simulation. Moreover, we find that the FRFPMS method inherently adds the contributions of phonon loss and gain within each frequency bin. Both of these issues related to the frequency-scaling can be fixed by a system-independent post-processing step.

\end{abstract}

\maketitle

\section{\label{sec:intro} Introduction}

Vibrational scattering is a rather new and very active field of research within Electron Energy Loss Spectroscopy (EELS) in Scanning Transmission Electron Microscopy (STEM) \cite{krivanek_vibrational_2014}, which can be roughly divided into damage-free aloof EELS and high spatial resolution vibrational EELS based on the following scattering mechanism: In aloof EELS, the electron scatters via an unscreened Coulomb interaction (``dipole scattering''), whereas scattering via the screened Coulomb interaction (``impact scattering'') gives rise to the high resolution component \cite{dwyer_electron-beam_2016}. Exploiting the unique properties of these components of vibrational EELS, many exciting experiments have been carried out, predominantly on systems of limited thickness and consisting of light atoms such as carbon, boron and nitrogen \cite{dwyer_electron-beam_2016,lagos_mapping_2017,lagos_thermometry_2018,hachtel_identification_2019,hage_phonon_2019,radtke_polarization_2019,senga_position_2019,collins_functional_2020,hage_single-atom_2020,xu_single-atom_2023}. For these systems, neglecting dynamical diffraction, i.e., the deformation of the elastic wave due to the strong elastic interaction between matter and beam electrons, may be a satisfactory approximation when describing the impact scattering component and such theories provide quite good agreement with experiment \cite{nicholls_theory_2019,hage_single-atom_2020} and have been used to study isotope effects in hexagonal Boron Nitride (hBN)-like molecules \cite{konecnaTheoryAtomicScaleVibrational2021}. Quantitative EELS at high spatial resolution on systems consisting of heavier atoms or being of larger thicknesses may, however, depend critically on the influence of dynamical diffraction and a more general theory of vibrational EELS should incorporate a proper treatment of elastic scattering. At the same time, the systems of most interest for vibrational EELS are systems of low or lowered symmetry, such as defects, disordered materials and (buried) nano-particles or nano-structures, where the high spatial and high energy resolution of the technique can be combined to enable unprecedented measurements of vibrational properties at the nano- and atomic-scale \cite{yan_single-defect_2021,cheng_experimental_2021,gadre_nanoscale_2022,hoglund_emergent_2022}. Therefore, a more general theory or model of vibrational EELS should also be able to deal with such systems, which require large super cells from a computational point of view.

There exist a few theories and simulation methods, which have been employed successfully to model phonon scattering in the impact scattering regime: pure single inelastic scattering theories \cite{nicholls_theory_2019,senga_position_2019,hage_single-atom_2020}, transition potential approaches \cite{allen_inelastic_1995,martin_model_2009,forbes_modeling_2016,dwyerProspectsSpatialResolution2017,rez_lattice_2021}, the Quantum Excitations of Phonons (QEP) model \cite{forbes_quantum_2010,lugg_atomic_2015} and the Frequency-Resolved Frozen Phonon Multislice (FRFPMS) method \cite{zeiger_efficient_2020,zeiger_frequency-resolved_2021}.
These theories originate from very different starting points and differ also in the rigor of their basic assumptions, which poses the obvious question of whether they lead to similar results and how they are related.
The single inelastic scattering theories are based on the approach of Van Hove \cite{van_hove_correlations_1954}, effectively a first-order Born approximation.
In the transition potential approaches, one can include the effects of beam shape and dynamical diffraction effects and the inelastic transition potential itself can be derived from different starting points, such as Yoshioka's coupled channel equations \cite{yoshioka_effect_1957,allen_inelastic_1995} or a more direct view involving the averaging of displaced scattering potentials \cite{amali_theory_1997}.
The QEP model on the other hand is derived from the full many-body Schrödinger equation of the combined system of the beam electron and the sample by applying an approximation similar in spirit to the well-known Born-Oppenheimer approximation to the beam electron wave function, but it does not immediately offer a computationally feasible access to the spectroscopic (energy-loss) dimension. Lastly the FRFPMS method is an intuitively motivated extension of the Frozen Phonon Multislice (FPMS) method \cite{loane_thermal_1991,muller_simulation_2001}, which allows spectral resolution, but was not formally derived or analytically investigated thus far.

The advantage of the FRFPMS method is that it is currently the only method which allows one to simulate vibrational spectra of large structure models of arbitrarily low symmetry, due to its linear scaling with system size. In order to illustrate this point in more detail, we consider how the computation of vibrational EELS would look in a single inelastic scattering formalism: one would first need to obtain all vibrational modes of the system in question (or a sufficiently dense sampling), meaning the vibrational frequencies $\omega_{\nu\mathbf{k}}$ and polarization vectors $\pmb{\epsilon}(\mathbf{k}\nu)$ for each phonon branch $\nu$ and crystal momentum $\mathbf{k}$. With this knowledge, one can assemble the transition potentials (c.f.\ Appendix~\ref{app:transition_potentials_Forbes}). A typical system model might be a $10 \times 10 \times 50$ supercell of SrTiO$_3$ with a lattice parameter of about $4$~\AA{} and 5 atoms in the unit cell. Such a supercell would be of dimensions $4\times 4\times 20$~nm$^3$ and contain $\mathcal{N} = 5 \times 10 \times 10 \times 50 = 25000$ atoms. Obtaining $\omega_{\nu\mathbf{k}}$ and $\pmb{\epsilon}(\mathbf{k}\nu)$ poses a formidable problem for a system of this size if the symmetry is low (consider having a defect within such a structure model), as it involves diagonalizing the so-called \emph{dynamical matrix} \cite{maradudin_theory_1963}, the size of which is $3\mathcal{N}\times3\mathcal{N}$. We note that the computational complexity of matrix diagonalization is less favourable than the square of the size of the matrix for many popular algorithms \cite{demmel_performance_2008}. But the difficulties with such an approach do not stop here, since one would also need to perform about $3\mathcal{N}$ independent elastic propagations of inelastic waves generated somewhere in the sample, further adding to the computational cost of such a simulation. Each of these propagations scales roughly with $\mathcal{N}$ itself if it is carried out by the multislice algorithm. The computational cost is further multiplied by the number of probe positions for cases which involve a scanning of the STEM probe. All these multiplicative factors render a full-sized transition potential calculation involving dynamical diffraction uneconomical with current computational resources.

In this article, we aim to relate single inelastic scattering theory, the QEP model and the FRFPMS method to each other for single inelastic phonon scattering. We show, using the example of an anharmonic quantum mechanical oscillator (AQHO), that in single inelastic scattering theory and the QEP model the inelastic phonon scattering cross section for a specific energy loss process $n_x \rightarrow  n_x+1$ is given by very similar expressions in the diffraction plane of the form
\begin{equation}
    \label{eq:inel_cross_section_intro}
    \frac{\langle n_x \rangle_T + 1}{\omega_x} \; \left({\mathbf{q} \pmb{\cdot} \pmb{\epsilon}_x }\right)^2 \; V_{\mathrm{proj}}^2 (q) \; e^{- 2 W(\mathbf{q})}.
\end{equation}
Here $\langle n_x \rangle_T$ is the thermally averaged occupancy of the assumed oscillator mode of frequency $\omega_x$, $q = |\mathbf{q}|$ the magnitude of the momentum transfer $\mathbf{q}=(q_x, q_y)$ and $V_{\mathrm{proj}}^2 (q) \; e^{- 2 W(\mathbf{q})}$ is the square of the projected potential smeared by an anisotropic Debye-Waller factor (DWF) $e^{- 2 W(\mathbf{q})}$. We show that a very similar expression to Eq.~\eqref{eq:inel_cross_section_intro} is obtained for the inelastically scattered intensity which results from applying the FRFPMS prescription to the scattered wave in the weak phase object approximation (WPOA)~\cite{kirkland_advanced_2010}, except for differences in the DWF and temperature- and energy-dependent scaling factor $\langle n_x \rangle_T + 1$. This result leads directly to a proposal for a revision of the FRFPMS method by smearing the scattering potential by the DWF instead of using bare scattering potentials, and a correction of the energy scaling. As a byproduct of our analytical considerations, we explain furthermore the $1/\omega^2$ energy-scaling of FRFPMS-EELS observed earlier in Ref.~\cite{zeiger_frequency-resolved_2021}. We also discuss the effect of temperature on EELS spectra and consider more generally the energy-scaling of phonon EELS. In this way, we reconcile the different approaches with each other.

The article is divided as follows: first, in Sec.~\ref{sec:parallels}, we introduce the three different theoretical approaches, derive within each approach the inelastic scattering cross section for an AQHO, point out parallels between the expressions and make suggestions on how to improve the FRFPMS method. In Sec.~\ref{sec:numerical_sims} we follow up on the suggested revision of the FRFPMS method and numerically investigate the elastically and inelastically scattered intensity as a function of momentum transfer. We illustrate thereby that such a \emph{revised} FRFPMS method reproduces the expected results in the form of the single-phonon inelastically scattered intensity as well as elastically scattered intensity, provided that the mean squared displacements (MSDs) in a frequency bin are a sufficiently small fraction of the total MSD. Lastly, in section~\ref{sec:conclusion_outlook}, we draw conclusions from this work and give an outlook on future directions of research.

\section{Parallels between different theoretical approaches to vibrational EELS}
\label{sec:parallels}

The purpose of this section is a comparison of the expressions, which arise in different theoretical approaches to vibrational EELS. In order to simplify the derivations, we consider a situation, in which a parallel electron beam is impinging on a single light atom, such that we can ignore dynamical diffraction effects to a good approximation. Under these conditions, both WPOA and Born approximation are applicable. We model the dynamics of the atom as a single 2D AQHO with Hamiltonian
\begin{align} \label{eq:aqho_hamiltonian}
H 
= & \frac{p_x^2}{2M} + \frac{p_y^2}{2M}  + \frac{M\omega_x^2 \tau_x^2}{2} + \frac{M\omega_y^2 \tau_y^2}{2}
\end{align}
where $\omega_x$ in general differs from $\omega_y$, $\mathbf{p} = (p_x,p_y)$ is the momentum operator, $\pmb{\tau} = (\tau_x,\tau_y)$ the displacement operator and $M$ the mass of the oscillator. The AQHO wave function can be written in terms of the harmonic oscillator wave functions for each dimension $a_{n_x}(\tau_x)$ and $a_{n_y}(\tau_y)$, i.e.,
\begin{equation}
    a_{\mathbf{n}}(\pmb{\tau}) = a_{n_x}(\tau_x) \, a_{n_y}(\tau_y),
\end{equation}
which satisfy
\begin{equation}
    H a_{\mathbf{n}}(\pmb{\tau}) 
    = \underbrace{\left[ \hbar\omega_x \left( \frac{1}{2} + n_x\right) +  \hbar\omega_y \left( \frac{1}{2} + n_y \right)\right]}_{= E_{\mathbf{n}}} a_{\mathbf{n}}(\pmb{\tau}),
\end{equation}
where $\mathbf{n} = (n_x, n_y)$ are the occupation numbers of the two energy eigenstates. Basic properties of a quantum harmonic oscillator, relevant for this work, are summarized in  Appendices~\ref{app:properties_qho_wf} and \ref{app:phonons_in_harm_approx}.

In the present work, we are interested in the scattering of high energy electrons travelling initially parallel to the $z$-axis. After interaction with the specimen of (very small) thickness $t$, the total wave function $\phi(\mathbf{r})$ of the electrons in the $xy$-plane, $\mathbf{r}=(x,y)$, can be represented in the WPOA as \cite{kirkland_advanced_2010}
\begin{equation}
    \phi(\mathbf{r}) = \left[1 + i\sigma V_{\mathrm{proj}}(\mathbf{r})\right] \phi_0(\mathbf{r}),
\end{equation}
where we neglect the exponential phase factor due to the propagation of the wave function along the $z$-direction. Furthermore $\sigma = m / (2\pi \hbar^2 k_0)$ is the interaction constant, $\phi_0(\mathbf{r})$ the incident wave, $m=\gamma m_e$ the relativistically corrected electron mass, $m_e$ the electron rest mass, $\gamma$ the Lorentz factor, $k_0$ is the relativistic wave number of the incident electron, and
\begin{equation}
    \label{eq:V_proj_definition}
    \begin{aligned}
        V_{\mathrm{proj}}(\mathbf{r}) 
        = {} & \int_{0}^{t} V(\mathbf{r},z) \mathrm{d}z \\
        = {} & \frac{2\pi\hbar^2}{m_e} \int_{0}^{t} \int f_e(\mathbf{Q}) \; e^{2\pi i\mathbf{Q}\pmb{\cdot}\mathbf{R}} \; \mathrm{d}\mathbf{Q} \mathrm{d}z
    \end{aligned}
\end{equation}
is the projected scattering potential of a single atom located at the origin of the coordinate system. Here we have defined $\mathbf{R} = (\mathbf{r},z)$, and $\mathbf{Q} = (\mathbf{q},q_z)$, where $\mathbf{q}=(q_x,q_y)$ is the momentum transfer in the $xy$-plane. The electron scattering factor is denoted by $f_e\left(\mathbf{Q}\right)$, which we assume to be isotropic in the remainder of this article, i.e., $f_e(\mathbf{Q}) = f_e(|\mathbf{Q}|) = f_e(Q)$. We further assume the incident wave to be of plane wave type $\phi_0(\mathbf{r}) = 1/L$, normalized to the area $L^2$ of a box with side length $L$, such that the Fourier transform of the total wave reads
\begin{equation}
    \label{eq:WPOA_reciprocal_space}
    \begin{aligned}
        \phi(\mathbf{q}) = \mathcal{FT}_{\mathbf{r}} \left[\phi(\mathbf{r})\right] (\mathbf{q})
        = {} & \frac{1}{L} \left[\delta(\mathbf{q}) + i\sigma V_{\mathrm{proj}}(\mathbf{q}) \right],
    \end{aligned}
\end{equation}
where $\mathcal{FT}_{\mathbf{r}} \left[\ldots\right] (\mathbf{q})$ denotes the 2D-Fourier transform and $V_{\mathrm{proj}}(\mathbf{q})$ is the 2D-Fourier transform of Eq.~\eqref{eq:V_proj_definition}. We assume the projection approximation, i.e.,
\begin{equation}
\begin{aligned}
    V_{\mathrm{proj}}(\mathbf{r}) 
    \approx {} & \int_{-\infty}^{\infty} V(\mathbf{r},z) \mathrm{d}z \\
    = {} & \frac{2\pi\hbar^2}{m_e} \int f_e(q) \, e^{2\pi i\mathbf{q}\pmb{\cdot}\mathbf{r}} \mathrm{d}\mathbf{q},
\end{aligned}
\end{equation}
where $q=|\mathbf{q}|$ and the Fourier transform of the projected potential
\begin{equation}
    V_{\mathrm{proj}}(q) = \frac{2\pi\hbar^2}{m_e} f_e(q)
\end{equation}
is then given in terms of the atomic electron scattering factor. In this situation the WPOA wave in reciprocal space simply becomes
\begin{equation}
    \label{eq:WPOA_reciprocal_space_1atom}
    \begin{aligned}
        \phi(\mathbf{q}) 
        = \frac{1}{L} \left[\delta(\mathbf{q}) + i\frac{\gamma}{k_0} f_e(q) \right]
    \end{aligned}
\end{equation}
and we will continue to use this expression in the text below.

For our considerations we need the connection between the scattering cross section and the intensity in WPOA. The second term in Eq.~\eqref{eq:WPOA_reciprocal_space} describes the scattered wave $\psi_{\mathrm{sc}}(\mathbf{q})$ and its squared absolute value is the scattered intensity towards momentum transfer $\mathbf{q}$. Multiplying this intensity with an infinitesimal area $k_0^2 \mathrm{d} \Omega_{\mathbf{q}}$ in reciprocal space gives the probability of scattering into the solid angle $\mathrm{d} \Omega_{\mathbf{q}}$. The cross section $\mathrm{d} \sigma$ associated with this scattered intensity is then related to the intensity of the scattered wave via
\begin{equation}
\label{eq:connection_2Dwave_function_to_cross_section}
    \frac{\mathrm{d} \sigma}{L^2}
    = {} \frac{\left| \phi_{\mathrm{sc}}(\mathbf{q})\right|^2 k_0^2 \; \mathrm{d} \Omega_{\mathbf{q}}}{I_0} 
\end{equation}
where $I_0 = \int |\phi_0(\mathbf{r})|^2 \mathrm{d}\mathbf{r} = 1$ is the initial intensity and $L^2$ the lateral size of the simulation box. The scattering cross section then becomes
\begin{equation}
    \label{eq:connection_cross_section_intensity_WPOA}
    \frac{\mathrm{d} \sigma }{\mathrm{d} \Omega_{\mathbf{q}}}
    = L^2 \frac{1}{L^2} \frac{\gamma^2}{k_0^2} f_e^2(q) k_0^2 
    = \gamma^2 f_e^2(q).
\end{equation}
Note that Eq.~\eqref{eq:connection_2Dwave_function_to_cross_section} is applicable to any elastically scattered wave known as a function of 2D momentum transfer $\mathbf{q}$. Specifically, it can also be used to relate a beam exit wave function from multislice simulations to the cross section. Note that for inelastic scattering, the scattered wave would travel at a different energy, which is not considered here, and we will see below, that the difference is practically negligible for phonon energy-losses in STEM.

We are interested in this section in the scattering probability for processes with energy loss $\Delta E = \hbar\omega_x$, i.e. $n_x \rightarrow n_x+1$ in the case of the AQHO. Section~\ref{sec:single_inel_scattering_theory} is dedicated to a discussion of these cases in the frameworks of single inelastic scattering theory and the transition potential formalism. Thereafter, in section~\ref{sec:QEP_in_WPOA}, we consider the QEP model and derive an expression for the inelastic scattering cross section for the cases described here. Finally, we derive an expression in WPOA for the FRFPMS method, compare it to the expressions derived within the other theories in section~\ref{sec:FRFPMS_in_WPOA}, and draw some conclusions about modifications of the FRFPMS method for more accurate simulations of single inelastic vibrational scattering.

\subsection{Single inelastic scattering theory}
\label{sec:single_inel_scattering_theory}

In the first order Born approximation, the double differential scattering cross section of single inelastic scattering with an energy loss ($\Delta E = \hbar\omega$) or gain ($\Delta E = -\hbar\omega$) and momentum transfer $\mathbf{Q} = \mathbf{Q}_0 + \mathbf{G}$, where $\mathbf{Q}_0$ is a vector in the 1-st Brillouin zone and $\mathbf{G}$ is a reciprocal lattice vector, can be expressed at temperature $T$ as \cite{nicholls_theory_2019,senga_position_2019,hage_single-atom_2020}
\begin{widetext}
\begin{equation}
\label{eq:cross_section_finiteT_single_inelastic_scattering}
\begin{aligned}
    \frac{\mathrm{d}^2 \sigma (\mathbf{Q},\omega,T) }{\mathrm{d} \Omega_{\mathbf{Q}} \mathrm{d} \omega} 
    = {} 2\pi^2 \hbar \frac{k_1}{k_0} \sum_{\nu} & \left| \sum_{j} \frac{1}{\sqrt{M_{j}}} e^{-2\pi i \mathbf{G} \pmb{\cdot}\mathbf{R}_{j}} \;  e^{- W_j(\mathbf{Q})} \; f_e^{(j)}(\mathbf{Q}) \; \mathbf{Q} \pmb{\cdot} \pmb{\epsilon}_{j}(\mathbf{Q}_0, \nu) \right|^2 \times \\
    & {} \times \left[ \frac{1+\langle n(\mathbf{Q}_0, \nu) \rangle_T}{\omega (\mathbf{Q}_0, \nu)} \; \delta(\omega-\omega (\mathbf{Q}_0, \nu)) + \frac{\langle n(\mathbf{Q}_0, \nu) \rangle_T}{\omega (\mathbf{Q}_0, \nu)} \; \delta(\omega + \omega (\mathbf{Q}_0, \nu))\right],
\end{aligned}
\end{equation}
\end{widetext}
where the first sum runs over phonon branches labeled by the index $\nu$ and the second sum over basis atoms labeled by the index $j$ at positions $\mathbf{R}_j$. $f_e^{j}(\mathbf{Q})$ is the electron scattering factor of the $j$-th basis atom as defined above, $\pmb{\epsilon}_{j}(\mathbf{Q}_0, \nu)$ is the phonon polarization vector of  atom $j$ in phonon mode ($\mathbf{Q}_0$,$\nu$).

The Dirac-$\delta$'s signify, that only excitations of modes with the correct energy $\hbar \omega (\mathbf{Q}_0, \nu)$ contribute to the cross section at a given energy transfer $\Delta E = \hbar \omega$. The $\delta(\omega - \omega (\mathbf{Q}_0, \nu))$ and $\delta(\omega + \omega (\mathbf{Q}_0, \nu))$ correspond then to the processes
\begin{subequations}
\begin{align}
    \text{energy loss:} \quad & n(\mathbf{Q}_0, \nu) \rightarrow n(\mathbf{Q}_0, \nu) + 1, \\
    \text{energy gain:} \quad & n(\mathbf{Q}_0, \nu) \rightarrow  n(\mathbf{Q}_0, \nu) - 1,
\end{align}
\end{subequations}
which refer to the emission and absorption of one quantum of energy by the beam electron, respectively. Overall these Dirac-$\delta$s give a non-trivial dependence of the inelastic cross section on energy-loss and momentum transfer. Furthermore the inelastic cross section is scaled in terms of energy by a global $1/\omega (\mathbf{Q}_0, \nu)$-factor, by the phonon occupation number $\langle n(\mathbf{Q}_0, \nu) \rangle_T = 1 / \left(e^{\beta \hbar \omega (\mathbf{Q}_0, \nu)} - 1 \right)$ at inverse temperature $\beta=1/k_{\mathrm{B}} T$, and the ratio $k_1/k_0$ of the scattered to the initial wave number. This ratio is the signature of the differing energies of the initial and scattered waves mentioned before, and for phonon scattering processes, this ratio is well approximated by unity: consider a typical phonon energy-loss of $\Delta E = 60$~meV, then the ratio becomes
\begin{equation*}
    \frac{k_1}{k_0} = \sqrt{1 - \frac{\Delta E}{E_0}} \approx 1 - \frac{1}{2} \frac{\Delta E}{E_0} = 1 - \frac{10^{-6}}{2}
\end{equation*}
for a typical initial energy $E_0=60$~keV. Here we have neglected the relativistic mass increase consistent with the non-relativistic nature of Eq.~\eqref{eq:cross_section_finiteT_single_inelastic_scattering}. In the remainder of this work, we will therefore set $k_1/k_0=1$.

Temperature has a two-fold effect in Eq.~\eqref{eq:cross_section_finiteT_single_inelastic_scattering}: on the one hand, a finite temperature leads to a scaling of the cross section due to the (thermally averaged) phonon occupation number $\langle n(\mathbf{Q}_0, \nu) \rangle_T$ and on the other hand there is an exponential damping of the atomic scattering factor at non-zero momentum transfers via the temperature-dependent DWF $e^{- 2 W_{j}(\mathbf{Q})}$ of the $j$-th atom. The DWF is related to the MSD of the $j$-th atom and thus includes contributions of all vibrational modes in the specimen. Appendix~\ref{app:DWF} contains further properties and definitions related to the DWF.

Beyond the already discussed energy and temperature dependencies encapsulated in the occupation number $\langle n(\mathbf{Q}_0, \nu) \rangle_T$, the inelastic cross section in Eq.~\eqref{eq:cross_section_finiteT_single_inelastic_scattering} contains non-trivial behavior as a function of the momentum transfer $\mathbf{Q}$ and as a function of mode frequency $\omega (\mathbf{Q}_0, \nu)$: the dot product of momentum transfer $\mathbf{Q}$ and phonon polarization vector $\pmb{\epsilon}_{j}(\mathbf{Q}_0, \nu)$ provides a directional selectivity. As a function of momentum transfer the absolute value of this dot product $\mathbf{Q} \pmb{\cdot} \pmb{\epsilon}_{j}(\mathbf{Q}_0, \nu)$ and the atomic scattering factor $f_e(\mathbf{Q})$ scale the inelastic scattering cross section.

We note here that a similar expression to equation \eqref{eq:cross_section_finiteT_single_inelastic_scattering} is considered in the form of a transition potential by Forbes et al. \cite{forbes_modeling_2016} for the situation of phonon scattering from the zero-phonon state $\left| \mathbf{0}\right\rangle$ (ground state) to an excited state $\left|\mathbf{n}\right\rangle$. Their theory is more general in that effects of beam shape can be included, but it does not consider the effects of initially thermally occupied modes (see also Eqns.~E.1 and D.3 in Ref.~\cite{martin_model_2009}), which requires an averaging of the inelastic cross section over all possible states at non-zero temperature. It is instructive to compare in detail the expression for the transition potential itself, which describes the inelastic scattering process in the approach of Forbes et al., to equation \eqref{eq:cross_section_finiteT_single_inelastic_scattering}. We show in appendix \ref{app:transition_potentials_Forbes} that in the single phonon scattering case both theories reduce to very similar expressions, matching when $\langle n(\mathbf{Q}_0, \nu) \rangle_T \ll 1$.%

\subsubsection{Anisotropic Quantum Harmonic Oscillator}
\label{sec:single_inel_scattering_theory_AQHO}

Consider now the case of a single atom modeled as a 2D AQHO with Hamiltonian given by Eq.~\eqref{eq:aqho_hamiltonian}. In this situation and with our assumptions of isotropic atomic scattering factors, Eq.~\eqref{eq:cross_section_finiteT_single_inelastic_scattering} reduces for the inelastic (energy loss) processes $n_x \rightarrow  n_x + 1$ to
\begin{equation}
    \label{eq:single_inel_cross_section_aniso_qmharm_osci}
    \begin{aligned}
        \frac{\mathrm{d}^2 \sigma (\mathbf{Q},\omega,T)}{\mathrm{d} \Omega_{\mathbf{q}} \mathrm{d} \omega} = & {} \frac{2\pi^2 \hbar}{M} \frac{1 + \langle n_x \rangle_T}{\omega_{x}} \; q_x^2 \; f_e^2(q)  \\
        & {} \times e^{-2W(\mathbf{q})} \delta(\omega - \omega_x),
    \end{aligned}
\end{equation}
where the DWF becomes (c.f.\ Appendix~\ref{app:DWF})
\begin{equation}
    \begin{aligned}
        e^{-2W(\mathbf{q})} 
        = {} & e^{-(2\pi)^2 \left\langle (\mathbf{q} \pmb{\cdot} \mathbf{u})^{2}\right\rangle_T} \\
        = {} & e^{(2\pi)^2 \left\langle (q_x u_x)^{2} \right\rangle_T} e^{(2\pi)^2 \left\langle (q_y u_y)^{2} \right\rangle_T} \\
        = {} & e^{-2W_x(q_x)} e^{-2W_y(q_y)},
    \end{aligned}
\end{equation}
since the displacements in the $x$ and $y$-direction are uncorrelated. The single inelastic scattering cross section in Eq.~\eqref{eq:single_inel_cross_section_aniso_qmharm_osci} exhibits an overall temperature and energy-dependent scaling by $(1+\langle n_x \rangle_T)/\omega_x$ and the $\mathbf{q}$-dependence is contained in the $q_x^2$ factor together with an electron scattering factor, which is thermally smeared by a DWF due to \emph{both} modes, with vibrations along $x$- as well as $y$-directions.

At $T=0$~K Eq.~\eqref{eq:single_inel_cross_section_aniso_qmharm_osci} reduces to
\begin{equation}
    \label{eq:single_inel_cross_section_aniso_qmharm_osci_T0K}
    \begin{aligned}
        \frac{\mathrm{d}^2 \sigma(\mathbf{q}, \omega, 0)}{\mathrm{d} \Omega_{\mathbf{q}} \mathrm{d} \omega}  = & {} \frac{2\pi^2 \hbar}{M} \frac{1}{\omega_{x}} \; q_x^2 \; f_e^2(q) e^{-2W_{0x}(q_x)}\\
        & {} \times e^{-2W_{0y}(q_y)} \; \delta(\omega - \omega_x),
    \end{aligned}
\end{equation}
where $W_{0x}(q_x)$ and $W_{0y}(q_y)$ are the DWFs associated with only the zero-point motion in the $x$- and $y$-direction, respectively. %

\subsection{QEP model in WPOA}
\label{sec:QEP_in_WPOA}

The QEP model \cite{forbes_quantum_2010,lugg_atomic_2015} is an alternative way to describe the phonon scattering process in a crystal: it is shown, that the full time-independent interacting many-body problem of beam electron and sample reduces to an effective time-independent Schrödinger equation for the beam electron, which is parametrically dependent upon the coordinates of all displaced atoms of the specimen. One assumes thereby a sufficient smallness of the variation of the beam electron wave function under infinitesimal changes in the nuclear coordinates \cite{lugg_atomic_2015}. This approximation is akin to the well-known Born-Oppenheimer approximation \cite{born_zur_1927}, under which the wave function of the electrons of a condensed matter system satisfy a time-independent Schrödinger equation, which is parametrically dependent upon the nuclear coordinates at every instance of time. The effective Schrödinger equation in the QEP model can be solved using the multislice method \cite{cowley_scattering_1957}.

In the QEP model the thermally averaged total intensity $\left\langle I (\mathbf{q}) \right\rangle_T$ in the diffraction plane is an incoherent average of so-called \emph{auxiliary wave functions} $\phi(\mathbf{q}, \pmb{\tau})$ over atomic configurations $\pmb{\tau}$ weighted by probability $P(\pmb{\tau})$ of a particular configuration $\pmb{\tau}$ being realized, i.e.,
\begin{equation}
    \label{eq:QEP_total_intensity}
    \left\langle I (\mathbf{q}) \right\rangle_T = \int \left| \phi(\mathbf{q}, \pmb{\tau}) \right|^2  P(\pmb{\tau} ) \; \mathrm{d}\pmb{\tau},
\end{equation}
where
\begin{align}
    \label{app_eq:correct_P(tau)_QEP}
    P(\pmb{\tau}) 
    = \left( \frac{1}{2\pi\langle \tau_{k}^2 \rangle} \right)^{\frac{3}{2} \mathcal{N}_{\mathrm{at}}} \; \prod_{k=1}^{3\mathcal{N}_{\mathrm{at}}} \exp\left[-\frac{\tau_{k}^2}{2\langle \tau_{k}^2 \rangle} \right]
\end{align}
is the joint probability distribution for an Einstein model of phonons in agreement with Ref.~\cite{lugg_atomic_2015}. We note that, due to a typo, a factor of -1/2 was accidentally omitted in the corresponding expression in Ref.~\cite{forbes_quantum_2010}. Note also, that for any practical purpose, the auxiliary waves $\phi(\mathbf{q}, \pmb{\tau})$ are the exit-plane wave functions obtained from a standard multislice calculation for the atomic configuration $\pmb{\tau}$.

The total intensity according to Eq.~\eqref{eq:QEP_total_intensity} can be thought of to be comprised of two additive contributions: the coherent intensity
\begin{equation}
    \label{eq:QEP_elastic_intensity}
    \left| \left\langle \psi (\mathbf{q})\right\rangle_T \right|^2 =  \left| \int \phi(\mathbf{q}, \pmb{\tau})  P(\pmb{\tau}) \, \mathrm{d}\pmb{\tau} \right|^2,
\end{equation}
which is the absolute square of a coherent average of auxiliary wave functions $\phi(\mathbf{q}, \pmb{\tau})$ over atomic configurations $\pmb{\tau}$, and the inelastic intensity
\begin{equation}
    \label{eq:QEP_inelastic_intensity}
    \left\langle I_{\mathrm{inel}} (\mathbf{q}) \right\rangle_T = \left\langle I (\mathbf{q}) \right\rangle_T - \left|\left\langle \psi (\mathbf{q}) \right\rangle_T\right|^2,
\end{equation}
which is effectively the variance of the auxiliary wave functions.

Lastly an ``inelastic wave`` associated with a transition $\left| \mathbf{m} \right\rangle \rightarrow \left| \mathbf{n} \right\rangle$ can be defined in the following way
\begin{equation}
    \label{eq:QEP_inelastic_wave_mn}
    \psi_{\mathbf{m}\mathbf{n}} (\mathbf{q}) = \int a_{\mathbf{n}}^*(\pmb{\tau} ) a_{\mathbf{m}}(\pmb{\tau}) \phi(\mathbf{q}, \pmb{\tau}) \; \mathrm{d}\pmb{\tau},
\end{equation}
where $a_{\mathbf{m}}(\pmb{\tau})$ and $a_{\mathbf{n}}^*(\pmb{\tau} )$ are the wave functions of the initial and final phonon states, respectively. This expression explicitly includes an arbitrary state $\mathbf{n}$ as the initial state and makes it thus suitable for thermal averaging over the initial state, which is in general not a pure state. Note that the meaning of $\mathbf{m}$ and $\mathbf{n}$ is changed compared to Refs.~\cite{forbes_quantum_2010,lugg_atomic_2015}. They consider a state "0" as the initial state, which is not necessarily the ground state, and their number $\mathbf{n}$ has a relative meaning in the sense of how many phonons are created or annihilated in the inelastic process. Here we use the ``occupation vectors'' $\mathbf{m}$ and $\mathbf{n}$ with an absolute meaning here, i.e., $\mathbf{n}$ is a vector of phonon occupation numbers.

The QEP model has been successfully used to simulate the total inelastic intensity associated with vibrational scattering using an Einstein model in a number of works \cite{forbes_quantum_2010,lugg_atomic_2015,hage_phonon_2019,hage_contrast_2020}. Lugg et al.\ showed furthermore that the analytical expression for the total intensity scattered inelastically on phonons in the QEP model matches the result derived in a transition potential formulation and scattering from the ground state $\mathbf{0}$ to an excited state $\mathbf{n}$ under arbitrary illumination conditions \cite{lugg_atomic_2015}. The inelastic waves given by Eq.~\eqref{eq:QEP_inelastic_wave_mn} have, however, not been explicitly calculated before.

We consider here the situation of the atom modeled as a 2D AQHO, which we described in the introduction to Sec.~\ref{sec:parallels}, see Eq.~\ref{eq:aqho_hamiltonian}. According to Eq.~\eqref{eq:QEP_inelastic_wave_mn} a single inelastic scattering event $n_x \rightarrow n_x+1$ creates an inelastic wave
\begin{equation}
    \label{eq:QEP__AQHO_psi_nn+1_integrals_basic}
    \psi_{n_x, n_x+1} (\mathbf{q}) = \int_{V} a_{(n_x+1,n_y)}^*(\pmb{\tau} ) a_{(n_x,n_y)}(\pmb{\tau}) \, \phi(\mathbf{q}, \pmb{\tau}) \; \mathrm{d}\pmb{\tau}.
\end{equation} 
$a_{(n_x,n_y)}(\pmb{\tau} ) = a_{n_x}(\tau_x) a_{n_y}(\tau_y)$ is the total (product) wave function of the wave functions $a_{n_x}(\tau_x)$ and $a_{n_y}(\tau_y)$ of the two harmonic oscillator energy eigenstates associated with the frequencies $\omega_{x}$ and $\omega_{y}$, respectively. We take the auxiliary wave $\phi(\mathbf{q}, \pmb{\tau})$ to be described in WPOA, i.e.,
\begin{equation}
    \label{eq:QEP_WPOA_wave}
    \phi(\mathbf{q}, \pmb{\tau}) 
    = \frac{1}{L} \left[\delta(q) + i \frac{\gamma}{k_0} f_e(q) e^{2\pi i q_x \tau_x} e^{2\pi i q_y \tau_y} \right],
\end{equation}
where we have represented the displaced potential in Fourier space and made use of the Fourier shift theorem. Before we consider the integral in Eq.~\eqref{eq:QEP__AQHO_psi_nn+1_integrals_basic} in detail, we need to connect the wave $\psi_{n_x, n_x+1} (\mathbf{q})$ to the inelastic cross section. To that end, we orient ourselves at Eq.~\eqref{eq:connection_2Dwave_function_to_cross_section}, so that we can write the cross section for the process $n_x \rightarrow n_x+1$ as
\begin{equation}
    \label{eq:QEP_cross_section_for_psi_n_n+1}
    \frac{\mathrm{d} \sigma_{n_x,n_x+1} (\mathbf{q})}{\mathrm{d} \Omega_{\mathbf{q}}\mathrm{d}\omega}
    = {} L^2 k_0^2 \left|\psi_{n_x, n_x+1}(\mathbf{q})\right|^2 \; \delta(\omega - \omega_x). 
\end{equation}
It should be noted, that the energy-conservation encapsulated in the Dirac-$\delta$ in this expression does not come out naturally from the QEP model, but needs to be enforced ``by hand'' at this stage. In order to compare this expression with the single inelastic scattering result in Eq.~\eqref{eq:single_inel_cross_section_aniso_qmharm_osci}, we need to perform a thermal average over the initial state $\mathbf{n}$ at temperature $T$ and enforce energy-conservation, i.e.,
\begin{equation}
    \label{eq:QEP_AQHO_ddscs_thermal_average_basic_expression}
        \frac{\mathrm{d}^2 \sigma (\mathbf{q}, \omega, T)}{\mathrm{d} \Omega_{\mathbf{q}} \mathrm{d} \omega}
        = \frac{1}{Z} \sum_{n_x,n_y} e^{-\beta E_{\mathbf{n}}} \frac{\mathrm{d} \sigma_{n_x, n_x+1} (\mathbf{q}, \omega)}{\mathrm{d} \Omega_{\mathbf{q}}},
\end{equation}
where $Z=\sum_{n_x,n_y} e^{-\beta E_{\mathbf{n}}}$ is the partition function.

Inserting Eq.~\eqref{eq:QEP_WPOA_wave} in Eq.~\eqref{eq:QEP__AQHO_psi_nn+1_integrals_basic} leads to the expression
\begin{widetext}
    \begin{equation}
    \begin{aligned}
        \label{eq:QEP_AQHO_psi_nn+1_integrals_an_an+1_eiqtau}
        \psi_{n_x, n_x+1}(\mathbf{q}) = {} & \frac{1}{L} \delta(q) \; \int a_{n_x+1}^*(\tau_x) a_{n_x}(\tau_x) \, d\tau_x \, \int a_{n_y}^*(\tau_y) a_{n_y}(\tau_y) \, d\tau_y \\
        {} + {} & {} \frac{1}{L} i\frac{\gamma}{k_0} f_e(q) \int a_{n_x+1}^*(\tau_x) a_{n_x}(\tau_x) \, e^{2\pi i q_x \tau_x} d\tau_x \, \int a_{n_y}^*(\tau_y) a_{n_y}(\tau_y) \, e^{2\pi i q_y \tau_y} d\tau_y.
    \end{aligned}
    \end{equation}
The first term vanishes due to the orthogonality of $a_{n_x+1}^*(\tau_x)$ and $a_{n_x}(\tau_x)$, see Appendix~\ref{app:properties_qho_wf}. The second term is of the same type as Eq.~\eqref{app_eq:int_e-x2_Hn_Hm_eibetax} and we show in Appendix~\ref{app:QEP_derivations_AQHO} how both of these integrals can be computed exactly (see also Ref.~\cite{martin_model_2009}). The result reads
    \begin{equation}
        \label{eq:QEP_AQHO_psi_nn+1_final}
            \psi_{n_x, n_x+1}(\mathbf{q}) 
            = - \frac{\pi \gamma}{k_0 L} \frac{\sqrt{2 M_x}}{\sqrt{n_x+1}} \; q_x f_e(q) \; e^{- W_{0x}(q_x)} e^{- W_{0y}(q_y)} \; L_{n_x}^1\big( 2 W_{0x}(q_x) \big) \; L_{n_y}^0 \big( 2 W_{0y}(q_y) \big),
    \end{equation}
\end{widetext}
where $L_{n}^{\alpha}$ are generalized Laguerre polynomials, and $M_x = \hbar / (M \omega_x)$.

For $n_x,n_y=0$, i.e., when the oscillator is initially in the ground state, we obtain ($L_{0}^{\alpha} = 1$ for any $\alpha$)
\begin{equation}
    \psi_{01}(\mathbf{q}) = - \frac{\pi \gamma}{k_0 L} \sqrt{2M_x} \; q_x f_e(q) \; e^{- W_{0x}(q_x)} e^{- W_{0y}(q_y)}
\end{equation}
and the cross section associated with this process of energy loss $\omega=\omega_x$ thus reads, according to Eq.~\eqref{eq:QEP_cross_section_for_psi_n_n+1}
\begin{equation}
    \label{eq:QEP_AQHO_ddscs_00+1_no_avg_general_init_state}
    \begin{aligned}
        \frac{\mathrm{d} \sigma_{0,1}(\mathbf{q})}{\mathrm{d} \Omega_{\mathbf{q}} \mathrm{d}\omega}
        = {} & \frac{2\pi^2\hbar \gamma^2}{M \omega_x} \; q_x^2 \; \delta(\omega - \omega_x) \\
        {} \times {} & f_e^2(q) \; e^{-2W_{0x}(q_x)} e^{-2 W_{0y}(q_y)} .
    \end{aligned}
\end{equation}
This expression is functionally the same as the expression for the inelastic scattering cross section in single inelastic Born approximation at $T=0$~K in Eq.~\eqref{eq:single_inel_cross_section_aniso_qmharm_osci_T0K}, except for the relativistic $\gamma$-factor.

We return to Eq.~\eqref{eq:QEP_AQHO_psi_nn+1_final} and consider the scattering cross section for $n_x\rightarrow n_x+1$ with an arbitrary initial state $\mathbf{n}=(n_x, n_y)$, i.e.,
\begin{equation}
\begin{aligned}
    \label{eq:QEP_AQHO_ddscs_nn+1_no_avg_general_init_state}
    \frac{\mathrm{d}^2 \sigma_{n_x, n_x+1}(\mathbf{q})}{\mathrm{d} \Omega_{\mathbf{q}}\mathrm{d}\omega}
    = {} & {} \pi^2 \gamma^2 \; q_x^2  \frac{2 M_x}{n_x+1} \; \delta(\omega - \omega_x) \\
    {} \times {} & {} \left\lbrace L_{n_x}^1\left[ 2 W_{0x}(q_x) \right] L_{n_y}^0\left[ 2 W_{0y}(q_y) \right] \right\rbrace^2 \\
    {} \times {} & {} f_e^2(q) \; e^{- 2W_{0x}(q_x)} e^{- 2W_{0y}(q_y)} .
\end{aligned}
\end{equation}
In order to compare this expression with the single inelastic scattering result in Eq.~\eqref{eq:single_inel_cross_section_aniso_qmharm_osci}, we need to perform the thermal average according to Eq.~\eqref{eq:QEP_AQHO_ddscs_thermal_average_basic_expression}. The details of the calculation can be found in Appendix~\ref{app:QEP_thermal_average_AQHO}. We quote here the result
\begin{eqnarray}
\label{eq:QEP_AQHO_ddscs_thermal_average_final}
    \lefteqn{\frac{\mathrm{d}^2 \sigma (\mathbf{q}, \omega, T)}{\mathrm{d} \Omega_{\mathbf{q}} \mathrm{d} \omega}} \nonumber \\
    & = & \gamma^2 \; f_e^2(q) \; e^{- 2W_{x}(q_x)} e^{- 2 W_{y}(q_y)} e^{\frac{\beta\hbar\omega_x}{2}} \; \delta(\omega-\omega_x) \nonumber \\ 
    & \times & I_1\left(\frac{2 W_{0x}(q_x)}{\sinh\left(\frac{\beta\hbar\omega_x}{2}\right)} \right) \; I_0\left(\frac{2 W_{0y}(q_y)}{\sinh\left(\frac{\beta\hbar\omega_y}{2}\right)} \right),
\end{eqnarray}
where $I_0$ and $I_1$ are modified Bessel functions. For small arguments $x \ll 1$, the modified Bessel functions behave as
\begin{equation}
    I_{m}(x) = \frac{1}{m!} \left( \frac{x}{2} \right)^m
\end{equation}
for any non-negative integer $m$. For a combination of sufficiently small momentum transfers $\mathbf{q}$, sufficiently low temperatures and sufficiently high frequencies $\omega_x$ and $\omega_y$, the arguments to the modified Bessel functions in Eq.~\eqref{eq:QEP_AQHO_ddscs_thermal_average_final} are in fact small  and we can approximate $I_0(x) \approx 1$ and $I_1 (x) \approx \frac{x}{2}$. We note that under these conditions the MSD of the corresponding mode is also small according to Eq.~\eqref{app_eq:properties_QHO_MSD}, i.e., the atom is not displaced far from its equilibrium position, and Eq.~\eqref{eq:QEP_AQHO_ddscs_thermal_average_final} becomes
\begin{equation}
    \label{eq:QEP_AQHO_ddscs_thermal_average_final_small_args_Bessel}
    \begin{aligned}
        \frac{\mathrm{d}^2 \sigma (\mathbf{q}, \omega, T)}{\mathrm{d} \Omega_{\mathbf{q}} \mathrm{d} \omega} {} = {} & {} \frac{2 \pi^2 \hbar \gamma^2}{M} \frac{\langle n_x \rangle_T + 1}{\omega_x} \; q_x^2 f_e^2(q) \\
          {} & {} \times e^{- 2 W_{x}(q_x)} e^{- 2W_{y}(q_y)} \delta(\omega-\omega_x),
    \end{aligned}
\end{equation}
where we have used
\begin{equation}
    \langle n_x \rangle_T + 1 = \frac{e^{\beta\hbar\omega_x/2}}{2\sinh(\beta\hbar\omega_x/2)}.
\end{equation}
Equation~\eqref{eq:QEP_AQHO_ddscs_thermal_average_final_small_args_Bessel} is the same expression as Eq.~\eqref{eq:single_inel_cross_section_aniso_qmharm_osci}, except for the relativistic $\gamma$-factor.

In a more general case, when the conditions of small momentum transfer $q$, low temperature $T$ and/or high mode frequency $\omega_x$ are not satisfied, the QEP model predicts the cross section in Eq.~\eqref{eq:QEP_AQHO_ddscs_thermal_average_final}. This expression involves the same thermally averaged scattering potential as Eqs.~\eqref{eq:QEP_AQHO_ddscs_thermal_average_final_small_args_Bessel} and \eqref{eq:single_inel_cross_section_aniso_qmharm_osci}, but the $q$-dependence of the remaining factors and the overall temperature scaling of the cross section is strikingly different. Modified Bessel functions appear instead of the simple $q^2$ and $(\langle n_x \rangle_T + 1)/\omega_x$ terms and it is not obvious, where this difference comes from. A deeper investigation of this issue is beyond the scope of the present work, but we believe that the answer lies in the treatment of the inelastic scattering potential, which is assumed to be \emph{small} in the case of the single inelastic scattering theory for the first-order Born approximation to be applicable. Such smallness of the inelastic potential is neither required not assumed in the QEP model, suggesting that the QEP result of Eq.~\eqref{eq:QEP_AQHO_ddscs_thermal_average_final} could be more general. Consistent with this line of thinking, the conditions of low temperature and/or high mode energy $\omega_x$, under which an argument of the modified Bessel functions becomes small for a certain mode, are effectively conditions under which the MSD of that same mode becomes small as well (c.f.\ Eq.~\eqref{app_eq:properties_QHO_MSD}). 

\subsection{FRFPMS method}
\label{sec:FRFPMS_in_WPOA}

The FRFPMS method has been introduced in Refs.~\cite{zeiger_efficient_2020,zeiger_frequency-resolved_2021}.
Here we briefly review the key equations and concepts of the method and then proceed to derive the inelastic intensity for the case of an AQHO, Eq.~\eqref{eq:aqho_hamiltonian}.

In the FRFPMS method we select a grid of $N_{\mathrm{bin}}$ frequencies $\omega_i$, $i=1,\ldots, N_{\mathrm{bin}}$ and for each of these frequencies run an FPMS simulation. In each of these FPMS simulations, the $N$ atomic structure snapshots are sampled in such a way that the displacements of atoms from their equilibrium positions in the structure correspond to the displacement contributions of modes within a certain range of frequencies around the selected frequency to the total atomic displacement (meaning that of all modes). This procedure is a priori independent of the way these snapshots are generated, but we have relied thus far on non-equilibrium MD simulations, in which a so-called $\delta$- or hotspot thermostat is used to supply energy to only a narrow range of frequencies in the simulation, while a strong damping keeps other modes ``frozen'' \cite{ceriotti_-thermostat:_2010,dettori_simulating_2017}.

In analogy to the QEP model outlined in Sec.~\ref{sec:QEP_in_WPOA}, the total, elastic and inelastic intensity in the diffraction plane are extracted by averaging the $N$ electron beam exit wave functions $\Psi\left(\mathbf{q}, \mathbf{r}_{\mathrm{b}}, \pmb{\tau}_n(\omega_i,T)\right)$ computed using the multislice algorithm for each energy bin $\omega_i$ and snapshot $\pmb{\tau}_n(\omega_i,T)$, where $n$ is the index of a snapshot and $\pmb{\tau}_n(\omega_i,T)$ itself contains displacements of all atoms in the structure model of snapshot $n$ within energy bin $i$ at temperature $T$:
\begin{subequations}
\label{eq:FRFPM_Iincoh_Icoh_Iinel}
\begin{align}
    I_{\mathrm{incoh}}(\mathbf{q}, \mathbf{r}_{\mathrm{b}}, \omega_i, T) 
    = {} & {} \frac{1}{N} \sum_{n=1}^{N} \left|\Psi\left(\mathbf{q}, \mathbf{r}_{\mathrm{b}}, \pmb{\tau}_n(\omega_i,T) \right) \right|^2 \nonumber\\
    = {} & {} \left\langle \left| \Psi\left(\mathbf{q}, \mathbf{r}_{\mathrm{b}}, \pmb{\tau}(\omega_i,T)\right) \right|^2 \right\rangle_{N} \label{eq:FRFPMS_Iincoh} \\
    I_{\mathrm{coh}}(\mathbf{q}, \mathbf{r}_{\mathrm{b}}, \omega_i) 
    = {} & {} \left|\frac{1}{N}  \sum_{n=1}^{N} \Psi\left(\mathbf{q}, \mathbf{r}_{\mathrm{b}}, \pmb{\tau}_n(\omega_i,T)\right) \right|^2 \nonumber\\
    = {} & {} \left| \left\langle \Psi \left(\mathbf{q}, \mathbf{r}_{\mathrm{b}}, \pmb{\tau}(\omega_i,T)\right) \right\rangle_{N} \right|^2 \label{eq:FRFPMS_Icoh} \\
    I_{\mathrm{vib}}(\mathbf{q}, \mathbf{r}_{\mathrm{b}}, \omega_i, T) 
    = {} & I_{\mathrm{incoh}}(\mathbf{q}, \mathbf{r}_{\mathrm{b}}, \omega_i,T) \nonumber\\
    {} & {} - I_{\mathrm{coh}}(\mathbf{q}, \mathbf{r}_{\mathrm{b}}, \omega_i,T) \nonumber\\
    = {} & \left\langle \left| \Psi\left(\mathbf{q}, \mathbf{r}_{\mathrm{b}}, \pmb{\tau}(\omega_i,T)\right) \right|^2 \right\rangle_{N} - \nonumber\\
      {} & - \left| \left\langle \Psi \left(\mathbf{q}, \mathbf{r}_{\mathrm{b}}, \pmb{\tau}(\omega_i,T)\right) \right\rangle_{N} \right|^2, \label{eq:FRFPMS_Ivib}
\end{align}
\end{subequations}
where $\mathbf{q}$ is the momentum transfer in the diffraction plane, $\mathbf{r}_{\mathrm{b}}$ the electron beam position in a STEM-EELS simulation, $I_{\mathrm{incoh}}$ is the incoherently averaged (i.e., total) intensity, $I_{\mathrm{coh}}$ the coherently averaged (i.e., elastic) intensity, and $I_{\mathrm{vib}}$ is the vibrational (inelastic) intensity, see analogous expressions within the QEP theory, Eqns.~\eqref{eq:QEP_total_intensity}, \eqref{eq:QEP_elastic_intensity} and \eqref{eq:QEP_inelastic_intensity}. We then define the inelastic double differential scattering cross section within the FRFPMS method to be
\begin{equation}
    \label{eq:FRFPMS_inel_cross_section_tot}
    \frac{\mathrm{d}^2 \sigma(\mathbf{q}, \mathbf{r}_{\mathrm{b}}, \omega, T)}{\mathrm{d} \Omega_{\mathbf{q}} \mathrm{d} \omega}  = \sum_{i=1}^{N_{\mathrm{bin}}} \delta(\omega - \omega_i) \; \frac{\mathrm{d} \sigma(\mathbf{q}, \mathbf{r}_{\mathrm{b}}, \omega_i,T)}{\mathrm{d} \Omega_{\mathbf{q}}},
\end{equation}
where
\begin{equation}
    \label{eq:FRFPMS_cross_section_I_vib}
    \frac{\mathrm{d} \sigma (\mathbf{q}, \mathbf{r}_{\mathrm{b}}, \omega_i, T)}{\mathrm{d} \Omega_{\mathbf{q}}}  = L^2 k_0^2 I_{\mathrm{vib}}(\mathbf{q}, \mathbf{r}_{\mathrm{b}}, \omega_i, T)
\end{equation}
is the cross section associated with the vibrational scattering in the $i$-th bin. For a plane incident wave, i.e., parallel illumination, which we consider in this work, we can drop the dependence on beam position $\mathbf{r}_{\mathrm{b}}$ in all of these expressions.

Thus we see, that the essence of the FRFPMS method is to let the atomic structure vibrate with a certain frequency $\omega$ and associate the inelastic intensity extracted from snapshots of this structure with the energy loss $\hbar\omega$. This idea can only be correct as long as multi-phonon processes are negligible, since if $I_{\mathrm{vib}}$ contained contributions due to multiple phonon or multi-phonon scattering, we would associate an incorrect energy loss with those processes. Furthermore, since we obtain a coherent wave for each frequency bin, and we attribute to this coherent wave the meaning of ``zero-loss'' wave or elastically scattered wave (including the direct beam), the coherent wave should be the same in each frequency bin, such that we can be sure, that the elastic wave, from which we generate the inelastic intensities $I_{\mathrm{vib}}$, has the correct thickness dependence. The effect of phonons on the elastic wave are two-fold: the averaged motion of nuclei leads to a DWF-smearing affecting dynamical diffraction, while the inelastic scattering itself leads to absorption. Thus the coherent wave associated with each bin, and from which the inelastic scattering is ``launched'', should have the same DWF-smearing and absorption in every bin. However, in this work, we will set aside the problem of absorption and focus purely on the DWF smearing in the coherent wave, which will be considered numerically in Sec.~\ref{sec:numerical_sims}.

\subsubsection{Anisotropic Quantum Harmonic Oscillator}
\label{sec:FRFPMS-aniso_qmharm_oscillator}

If we apply the ideas of the FRFPMS method to our model situation of an AQHO, we would pick two frequency bins, centered around $\omega_{x}$ and $\omega_{y}$, respectively, for which we need to sample atomic structure snapshots. These snapshots contain only displacements in the $x$- or $y$-direction, which are consistent with the mean-square displacement $\left\langle \tau_x^{2} \right\rangle_T$ or $\left\langle \tau_y^{2} \right\rangle_T$ for a frequency of $\omega_{x}$ or $\omega_{y}$, respectively.

We now focus our attention on the intensities calculated for the $\omega_x$ frequency bin. The instantaneous potential in an atomic structure snapshot reads
\begin{equation}
    V_{\mathrm{proj}}(\mathbf{q}, \tau_x) = V_{\mathrm{proj}}(q) \; e^{2\pi i q_x \tau_x},
\end{equation}
where $\tau_x$ is the displacement of the AQHO from the origin and $V_{\mathrm{proj}}(q)$ is the non-displaced projected potential.

We show in detail in Appendix~\ref{app:FRFPMS_WPOA}, how the different averages prescribed by Eqns.~\eqref{eq:FRFPMS_Iincoh}, \eqref{eq:FRFPMS_Icoh}, and \eqref{eq:FRFPMS_Ivib} are evaluated in the present situation. We quote here the main results. The incoherent intensity reads
\begin{equation}
    \label{eq:FRFPMS_AQHO_Iincoh}
    \begin{aligned}
        I_{\mathrm{incoh}} (\mathbf{q}, \omega_x, T) 
        = {} & \left\langle \left|\phi(\mathbf{q}, \omega_x, T) \right|^2 \right\rangle \\
        = {} & \frac{1}{L^2} \left[ \delta(q) + \frac{\gamma^2}{k_0^2} f_e^2(q) \right].
    \end{aligned}
\end{equation}
At a first glance it may seem counter-intuitive that there is no DWF in this expression. However, in our AQHO model one illuminates a single atom by a plane-wave electron beam. Therefore displacements of the atom do not play any role in an \emph{incoherent} average, since the plane incident wave is not sensitive to position and the total scattered intensity remains the same as for an atom at rest. For the coherent average one obtains
\begin{equation}
    \label{eq:FRFPMS_AQHO_Icoh}
    \begin{aligned}
        I_{\mathrm{coh}} (\mathbf{q}, \omega_x, T)
        = {} & \left|\left\langle \phi(\mathbf{q}, \omega_x, T) \right\rangle \right|^2 \\
        = {} & \frac{1}{L^2} \left[ \delta(q) + \frac{\gamma^2}{k_0^2} f_e^2(q) e^{-2W_x(q_x)} \right],
    \end{aligned}
\end{equation}
where the exponential term provides a Debye-Waller type smearing with $W_x(q_x) = 2\pi^2 q_x^2 \left\langle \tau_x^{2} \right\rangle_T$, however, only due to displacements along the $x$-direction. We return to this important point below. Using Eqns.~\eqref{eq:FRFPMS_AQHO_Iincoh} and \eqref{eq:FRFPMS_AQHO_Icoh}, the inelastic part reads
\begin{equation}
    \label{eq:FRFPMS_AQHO_Ivib}
    \begin{aligned}
        I_{\mathrm{vib}} (\mathbf{q}, \omega_x, T)
        = {} & I_{\mathrm{incoh}} (\mathbf{q}, \omega_x, T) - I_{\mathrm{coh}} (\mathbf{q}, \omega_x, T) \\
        = {} &  \frac{\gamma^2}{k_0^2 L^2} f_e^2(q) \left[ 1 - e^{-2W_x(q_x)} \right] \\
        = {} & \frac{\gamma^2}{k_0^2 L^2} f_e^2(q) \sum_{n=1}^{\infty} \frac{(-1)^{n+1} \left(4 \pi^2 q_x^2 \left\langle \tau_x^{2} \right\rangle_T \right)^n}{n!}.
    \end{aligned}
\end{equation}
In the derivation of Eq.~\eqref{eq:QEP_AQHO_ddscs_thermal_average_final_small_args_Bessel} small scattering angles (momentum transfers) and a small MSD were assumed. Under similar conditions, we retain only the $n=1$ term of Eq.~\eqref{eq:FRFPMS_AQHO_Ivib} and, according to Eqs.~\eqref{eq:FRFPMS_inel_cross_section_tot} and \eqref{eq:FRFPMS_cross_section_I_vib}, the inelastic cross section then becomes
\begin{align}
    \label{eq:FRFPMS_Ivib_final}
    \begin{aligned}
        \frac{\mathrm{d}^2 \sigma (\mathbf{q}, \omega, T)}{\mathrm{d} \Omega_{\mathbf{q}} \mathrm{d} \omega}  
        = {} & {} 4 \pi^2 \gamma^2 q_x^2 \left\langle \tau_x^{2} \right\rangle_T f_e^2(q) \; \delta(\omega-\omega_x)
    \end{aligned}
\end{align}
for the inelastic scattering on the $\omega_x$-mode. The cross section in the FRFPMS model has a similar $q^2$-momentum transfer dependence as in Eqs.~\eqref{eq:single_inel_cross_section_aniso_qmharm_osci} and \eqref{eq:QEP_AQHO_ddscs_thermal_average_final_small_args_Bessel}, however, there is a significant difference in the lack of full DWF smearing, which leads to an overestimation of the inelastic scattering to large angles, which are beyond the applied approximation. In Sec.~\ref{sec:FRFPMS_improvements} we show how this can be simply and effectively remedied.

The other difference is the presence of the MSD $\left\langle \tau_x^{2} \right\rangle_T$ along the $x$-direction in the FRFPMS results instead of a term proportional to $(\langle n_x \rangle_T + 1)/\omega_x$. In the quantum-mechanical statistics, the MSD within the $x$-dimension of the AQHO reads (see Eq.~\eqref{app_eq:properties_QHO_MSD})
\begin{equation}
    \label{eq:FRFPMS_MSD_qm_factor2}
    \begin{aligned}
        \langle \tau_x^2 \rangle_T 
        = {} & \frac{\hbar}{2M\omega_x} \coth \frac{\beta \hbar\omega_x}{2} \\
        = {} & \frac{\hbar}{2M\omega_x} \left[\frac{2}{e^{\beta \hbar\omega_x} - 1} + 1 \right] \\
        = {} & \frac{\hbar}{2M\omega_x} \left[2 \langle n_x \rangle_T + 1 \right],
    \end{aligned}
\end{equation}
so that the difference between the MSD and the factor $(\langle n_x \rangle_T + 1) / \omega_x$ lies in the factor of two in front of the phonon occupation number. Thus, whenever $\langle n_x \rangle_T$ is non-negligible, this would lead to a different energy and temperature scaling of the inelastic cross section than in the other approaches. While this affects only the scaling of spectra and not the position of peaks, the scaling is non-uniform throughout the range of energy losses, which has consequences for matching peak heights and shapes to experiment. It is important to note, though, that this scaling is a universal (i.e., system-independent) function of frequency and can be corrected for, as is outlined in Sec.~\ref{sec:FRFPMS_improvements}.

\begin{figure}
    \centering
    \includegraphics[width=\linewidth]{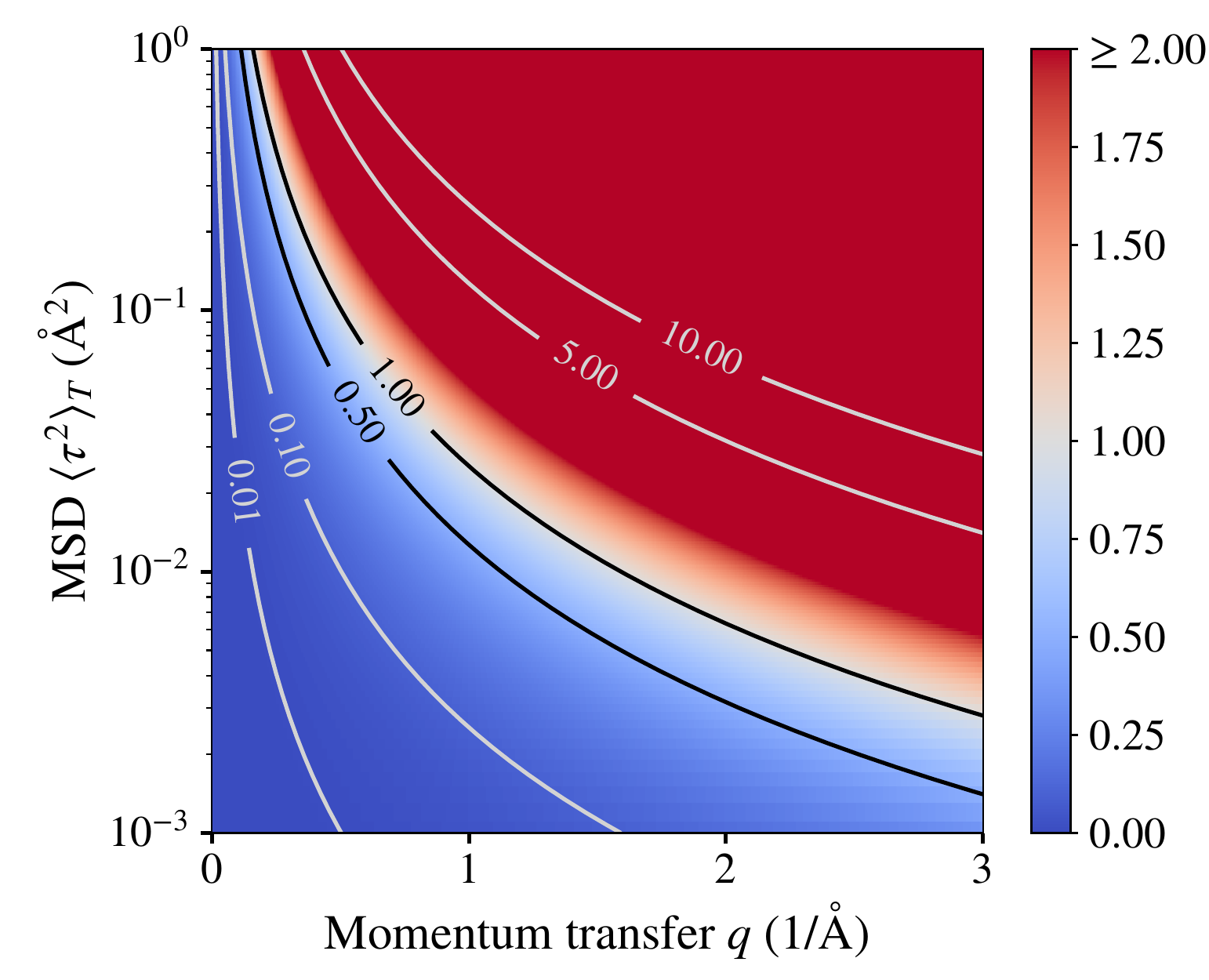}
    \caption{Color map of the exponent of the DWF $2W(q) = 2 \pi^2 q^2 \left\langle \pmb{\tau}^{2} \right\rangle_T$ as a function of the 2D MSD $\left\langle \pmb{\tau}^{2} \right\rangle_T$ and the momentum transfer $q$. Compare to the requirement in Eq.~\eqref{eq:FRFPMS_AQHO_1st_order_requirement}.}
    \label{fig:Figures/plot_1st_order_requirement}
\end{figure}

An apparent question, which arises from the considerations in this section, is under which conditions we can retain only the first order in Eq.~\eqref{eq:FRFPMS_AQHO_Ivib}. The answer is, that
\begin{equation}
    \label{eq:FRFPMS_AQHO_1st_order_requirement}
    2 W(q_x) = 4 \pi^2 q_x^2 \left\langle \tau_x^{2} \right\rangle_T \overset{!}{\ll} 1,
\end{equation}
i.e., we require, that the DWF exponent is sufficiently small. We plot the value of the right hand side of this relation in Fig.~\ref{fig:Figures/plot_1st_order_requirement} as a function of $q_x$ and $\left\langle \tau_x^{2} \right\rangle_T$. Typical values of the MSD lie in the range of about 0.005 to 0.5~\AA$^2$ for cubic elemental crystals at room temperature \cite{butt_compilation_1988}, so we can expect, that the expansion is a reliable approximation for many materials at momentum transfers below about 1~\AA$^{-1}$ such that $2W(q_x) < 0.1$. We will verify through numerical simulation in Sec.~\ref{sec:numerical_sims}, that this criterion is sufficient.

Note that the same expression as Eq.~\ref{eq:FRFPMS_Ivib_final} can be obtained also using real-space considerations. Assuming that the displacements $\bm{\tau}$ of the atom are small, the shifted potential can be well approximated by a Taylor expansion to the first order as $V(\mathbf{r} + \bm{\tau}) \approx V(\mathbf{r}) + \bm{\tau} \cdot \nabla V(\mathbf{r})$. The first term is responsible for scattering as if on a static atom, while the displacement-dependent part appears due to atomic vibrations. In real-space, the WPOA would lead to a displacement-dependent modification of the electron beam wave-function $\delta\psi(\mathbf{r}) = i\sigma \bm{\tau} \cdot \nabla V(\mathbf{r})$, which in Fourier space reads $\delta\psi(\mathbf{q}) = -2\pi \sigma \bm{\tau} \cdot \mathbf{q} V(\mathbf{q})$, see Eq.~\eqref{app_eq:fourier_transform_gradient}. Considering only displacements along the $x$-direction, taking the amplitude squared and incoherent averaging over the resulting squared displacements $\tau_x$ we recover
\begin{equation}
    \begin{aligned}
        \langle |\delta\psi(\mathbf{q})|^2 \rangle_T 
        = {} & 4\pi^2 \sigma^2 \langle \tau_x^2 \rangle_T q_x^2 V(q)^2 \\
        = {} & 4\pi^2 \frac{\gamma^2}{k_0^2} q_x^2 \langle \tau_x^2 \rangle_T f_e^2(q)
    \end{aligned}
\end{equation}
consistent with Eq.~\ref{eq:FRFPMS_Ivib_final}.

\subsubsection{Avenues of improvement of the FRFPMS method}
\label{sec:FRFPMS_improvements}

The goal with the FRFPMS method is to model single phonon energy-loss scattering for large systems without direct knowledge of all phonon modes. In this section we discuss ideas of how to improve the \emph{original} {FRFPMS} method \cite{zeiger_efficient_2020,zeiger_frequency-resolved_2021}, which are motivated by the preceding considerations of phonon scattering in different theories. To this end, we find by comparing the expressions for single inelastic scattering, Eqns.~\eqref{eq:single_inel_cross_section_aniso_qmharm_osci} and \eqref{eq:QEP_AQHO_ddscs_thermal_average_final_small_args_Bessel} to Eq.~\eqref{eq:FRFPMS_Ivib_final}, that instead of the bare potential a DWF-smearing should be included in FRFPMS calculations, which takes into account the smearing of the potential due to all thermally excited phonon modes in the system. Effectively, that means, that we would modify the scattering potential according to
\begin{equation}
    \label{eq:FRFPMS_correction_DWF}
    \begin{aligned}
        V_{\mathrm{proj}}(\mathbf{q}, \tau_x) \rightarrow {} & V_{\mathrm{proj}}(\mathbf{q}, \tau_x) \; e^{-W(\mathbf{q})} \\
        = {} & V_{\mathrm{proj}}(q) \; e^{2\pi i q_x \tau_x} \; e^{-W(\mathbf{q})}.
    \end{aligned}
\end{equation}
The smearing by a Debye-Waller factor in Eq.~\eqref{eq:FRFPMS_correction_DWF} will exponentially suppress high-angle scattering, both elastic as well as inelastic. In turn the resulting inelastic intensity in such a \emph{revised} FRFPMS should approximate the momentum transfer dependence of single inelastic scattering better and also elastic scattering should be modeled much more accurately by improving the period of so-called \emph{Pendellösung} oscillations of dynamical diffraction. We note here that the DWF is in general anisotropic for many models. However, this is rarely taken into account in the (S)TEM literature \cite{weickenmeier_influence_1998}. We suggest therefore, that it is sufficient to use isotropic DWFs as a first step in improving FRFPMS simulation and we consider exactly this scenario in the simulations in Section~\ref{sec:numerical_sims} and leave the influence of anisotropic MSDs open for future investigations.

We have seen that the MSD of the vibrational mode in question appears in Eq.~\eqref{eq:FRFPMS_Ivib_final} instead of the factor $(\langle n_x \rangle_T + 1)/\omega_x$. The difference between the two is a factor of ``2'' next to the $\langle n_x \rangle$, as discussed in connection with Eq.~\eqref{eq:FRFPMS_MSD_qm_factor2}. We propose here, that the FRFPMS method as we have used it actually adds the intensities corresponding to energy-loss and -gain, since their respective probabilities are proportional to $\langle n_x \rangle_T + 1$ and $\langle n_x \rangle_T$. We have verified in a separate, but analogous calculation to appendices~\ref{app:QEP_derivations_AQHO} and \ref{app:QEP_thermal_average_AQHO}, that the process $n_x\rightarrow n_x-1$ yields the same as Eq.~\eqref{eq:QEP_AQHO_ddscs_thermal_average_final_small_args_Bessel}, except that the factor $\langle n_x \rangle_T + 1$ becomes $\langle n_x \rangle_T$. We note furthermore, that this interpretation suggests furthermore, that both the FPMS method and also the incoherent and total inelastically scattered intensities in the QEP model according to Eqs.~\eqref{eq:QEP_total_intensity} and \eqref{eq:QEP_inelastic_intensity}, respectively, take both energy-loss as well as energy-gain processes into account.

Now that we have an understanding of the origin of the $2\langle n_x \rangle + 1$-factor, we can focus our attention on how to correct the FRFPMS expression for single inelastic energy-loss scattering: one might devise new sampling methods for frequency-resolved snapshots, in which the snapshots exhibit a rescaled MSD, that complies with the $\langle n_x \rangle_T + 1$ behavior. A simpler approach, which would also allow one to model the energy-gain side of the cross section, would be to rescale the inelastic signal in each frequency bin when assembling the cross section, i.e., we propose to replace the prescription of Eq.~\eqref{eq:FRFPMS_inel_cross_section_tot} by
\begin{widetext}
\begin{equation}
    \label{eq:improvements_FRFPMS_spectral_rescaling}
    \frac{\mathrm{d}^2 \sigma(\mathbf{q}, \mathbf{r}_{\mathrm{b}}, \omega, T)}{\mathrm{d} \Omega_{\mathbf{q}} \mathrm{d} \omega} \\
    = \sum_{i=1}^{N_{\mathrm{bin}}} \left[\frac{\langle n_x(\omega_i)\rangle_T+1}{2\langle n_x(\omega_i)\rangle_T+1} \; \frac{\mathrm{d} \sigma(\mathbf{q}, \mathbf{r}_{\mathrm{b}}, \omega_i,T)}{\mathrm{d} \Omega_{\mathbf{q}}} \delta(\omega - \omega_i) + \frac{\langle n_x(\omega_i)\rangle_T}{2\langle n_x(\omega_i)\rangle_T+1}  \; \frac{\mathrm{d} \sigma(\mathbf{q}, \mathbf{r}_{\mathrm{b}}, \omega_i,T)}{\mathrm{d} \Omega_{\mathbf{q}}}\delta(\omega + \omega_i) \right],
\end{equation}
\end{widetext}
where $\langle n_x(\omega_i)\rangle_T = 1 / \left(e^{\beta \hbar \omega_i} - 1 \right)$ is the occupation number representative of the frequency bin associated with frequency $\omega_i$. The rescaling prescribed by Eq.~\eqref{eq:improvements_FRFPMS_spectral_rescaling} assumes that the MSD in FRFPMS spectra follows quantum statistics, which has not been the case in our previous works in Refs.~\cite{zeiger_efficient_2020,zeiger_frequency-resolved_2021,zeiger_simulations_2021}. We discuss this point and the consequences in more detail in section~\ref{sec:FRFPMS_energy_scaling} below.

\subsubsection{Energy scaling of vibrational EELS spectra}
\label{sec:FRFPMS_energy_scaling}

In this section we discuss the energy-scaling expected for vibrational EELS spectra, as this has led to some confusion in discussions related to our previous work in Ref.~\cite{zeiger_frequency-resolved_2021}. The reason for the confusion is that from Eq.~\eqref{eq:cross_section_finiteT_single_inelastic_scattering} and even more so from Eq.~\eqref{app_eq:forbes_et_al_Single_excitation_cross_section}, one is naively tempted to expect that the cross section scales as $1/\omega$. Following this chain of thought and arguments outlined in the supplementary material to Ref.~\cite{hage_single-atom_2020}, one could expect that the intensity $I(\omega)$ measured at a certain energy loss $\hbar\omega$ for a large enough off-axis detector $\Omega_{\mathbf{q}_0}$ multiplied by $\omega$ is comparable to the phonon density of states (PDOS) $g(\omega)$, i.e.
\begin{equation}
   \omega \; I(\omega) = \omega \int_{\Omega_{\mathbf{q}_0}} \, \frac{\mathrm{d}^2 \sigma}{\mathrm{d} \Omega_{\mathbf{q}} \mathrm{d} \omega} (\mathbf{q}, \omega) d\mathbf{q} \propto  g (\omega).
\end{equation}
We observed in Ref.~\cite{zeiger_frequency-resolved_2021}, however, that in our FRFPMS calculation the quantity $\omega^2 \, I(\omega)$ compared much better with the PDOS. As it turns out, there are several issues, which need to be disentangled here.

We start by considering the expression for the double differential scattering cross section in single inelastic Born approximation scattering theory, Eq.~\eqref{eq:cross_section_finiteT_single_inelastic_scattering}. We have seen that the main temperature and energy-dependence is encapsulated in a factor $(\langle n \rangle_T + 1)/\omega$, which has the following limiting behavior (note that we multiply by a factor of $\hbar/(2M)$ in order to get comparable units with later expressions)
\begin{equation}
    \label{eq:frfpms_energy_scaling_n+1_limits}
    \begin{aligned}
    \frac{\hbar}{2M} \frac{\langle n \rangle_T + 1}{\omega} 
    = {} & \frac{\hbar}{2M\omega} \left[\frac{1}{e^{\beta\hbar \omega} - 1} + 1 \right] \\
    \approx {} & 
    \begin{cases}
         \displaystyle\frac{1}{2\beta\omega^2 M}   & \mathrm{for}~ \beta\hbar\omega \rightarrow 0 \\
         \displaystyle\frac{\hbar}{2\omega M}     & \mathrm{for}~ \beta\hbar\omega \rightarrow \infty
    \end{cases}.
    \end{aligned}
\end{equation}
Thus, for large enough energy losses in comparison with $\beta^{-1} = k_{\mathrm{B}}T$, we obtain a scaling of $1/\omega$, and $1/\omega^2$ for small energy losses compared with $\beta^{-1}$. This means, that neither scaling is valid for the entire spectral range, contrary to the first naive expectation outlined previously. Nevertheless, many calculations performed thus far have considered basically a temperature $T=0$~K, as we show for example in Appendix~\ref{app:transition_potentials_Forbes}, and these theories scale therefore as $1/\omega$. We also note here, that the energy loss $\hbar\tilde{\omega}$, for which $\beta\hbar \tilde{\omega} = 1$ is about $26$~meV at 300~K. This energy $\hbar\tilde{\omega}$ gives a criterion for which energy ranges the limits in Eq.~\eqref{eq:frfpms_energy_scaling_n+1_limits} are applicable: for $\omega \ll \tilde{\omega}$ we expect the phonon EELS to scale as $1/\omega^2$, and for $\omega \gg \tilde{\omega}$ to scale as $1/\omega$.

After establishing our expectations from careful consideration of single inelastic scattering theory, we can turn to the question of the energy scaling of FRFPMS spectra. We have seen, that the FRFPMS cross section in Eq.~\eqref{eq:FRFPMS_Ivib_final} scales with the MSD of the phonon mode which is to be excited. We pointed out in the discussion of this expression that the MSD scales as $(2\langle n \rangle_T +1)/\omega$. Therefore the limiting behavior of the cross section in Eq.~\eqref{eq:FRFPMS_Ivib_final} for small and large energies reads
\begin{equation}
    \label{eq:frfpms_energy_scaling_2n+1_limits}
    \begin{aligned}
        \left\langle \tau^2 \right\rangle_T 
        = {} & \frac{\hbar}{2M} \frac{2\langle n \rangle_T + 1}{\omega}
        = \frac{\hbar}{2M\omega} \coth \frac{\beta \hbar\omega}{2} \\
        \approx {} & 
        \begin{cases}
            \displaystyle\frac{1}{\beta \omega^2 M}  & \mathrm{for}~ \beta\hbar\omega \rightarrow 0 \\
            \displaystyle\frac{\hbar}{2\omega M}     & \mathrm{for}~ \beta\hbar\omega \rightarrow \infty
        \end{cases}.
    \end{aligned}
\end{equation}
At large enough energy losses in comparison with $\beta$, we thus obtain a scaling of $1/\omega$, whereas at small energy-losses in comparison with $\beta$, the cross section scales as $1/\omega^2$, albeit with a factor of ``2'' missing in comparison with the scaling behavior in Eq.~\eqref{eq:frfpms_energy_scaling_n+1_limits}, which is the signature of the implicit addition of energy-gain and energy-loss processes as discussed in section~\ref{sec:FRFPMS_improvements}.

In the FRFPMS method as we used it thus far, we sample modes with classical statistics and the MSD of a 1D classical harmonic oscillator scales as
\begin{equation}
    {\left\langle \tau^2 \right\rangle_T}^{(\mathrm{cl})} = \frac{1}{\beta \omega^2 M},
\end{equation}
which can be quickly inferred by taking the $\hbar\rightarrow 0$ limit of Eq.~\eqref{eq:frfpms_energy_scaling_2n+1_limits}. Thus the observation we made about the closer correspondence of the EELS multiplied by the square of the energy, i.e., $\omega^2 I(\omega)$, with the PDOS in Ref.~\cite{zeiger_frequency-resolved_2021} can be explained by the classical statistics of the MSD in the snapshots, since the multiplication of the spectrum by $\omega^2$ will exactly cancel the scaling contributed by the classical MSD. We have already outlined generally in section~\ref{sec:FRFPMS_improvements}, that we can improve the FRFPMS description by rescaling the spectra akin to Eq.~\eqref{eq:improvements_FRFPMS_spectral_rescaling}, but in the case that the MSD follows classical statistics, said procedure should rather read
\begin{widetext}
    \begin{equation}
    \frac{\mathrm{d}^2 \sigma(\mathbf{q}, \mathbf{r}_{\mathrm{b}}, \omega, T)}{\mathrm{d} \Omega_{\mathbf{q}} \mathrm{d} \omega}
    = \sum_{i=1}^{N_{\mathrm{bin}}} \frac{ \beta\hbar\omega_i }{2} \left\lbrace \left[\langle n_x(\omega_i)\rangle_T+1\right] \; \delta(\omega - \omega_i) \; \frac{\mathrm{d} \sigma(\mathbf{q}, \mathbf{r}_{\mathrm{b}}, \omega_i,T)}{\mathrm{d} \Omega_{\mathbf{q}}} + \langle n_x(\omega_i)\rangle_T \; \delta(\omega + \omega_i) \; \frac{\mathrm{d} \sigma(\mathbf{q}, \mathbf{r}_{\mathrm{b}}, \omega_i,T)}{\mathrm{d} \Omega_{\mathbf{q}}} \right\rbrace,
    \label{eq:improvements_FRFPMS_spectral_rescaling_classic_MSD}
\end{equation}
\end{widetext}
where $\frac{\mathrm{d} \sigma(\mathbf{q}, \mathbf{r}_{\mathrm{b}}, \omega_i,T)}{\mathrm{d} \Omega_{\mathbf{q}}}$ is the inelastic cross section obtained from the classically sampled snapshots.

\section{Numerical simulations}
\label{sec:numerical_sims}

In this section, we show numerically, using multislice calculations \cite{cowley_scattering_1957}, that the conclusions we have drawn previously about modifying the prescription for the FRFPMS by inclusion of a DWF-smeared potential, gives an improved description of the inelastically scattered intensity as well as the coherent intensity (sum of the intensities of the unscattered direct beam and elastically scattered electrons). For simplicity of the calculations, we consider a target consisting of a single carbon atom, whose vibrations are those of a 2D isotropic quantum harmonic oscillator (IQHO), contrary to section~\ref{sec:parallels}, in which we have considered an AQHO. The IQHO is, however, the limit $\omega_y \rightarrow \omega_x = \omega$ of the AQHO.

Specifically, we take the MSD of the 2D IQHO $\langle \pmb{\tau}^2 \rangle_T$ to have a value of $\mathrm{0.01}~\mathrm{nm}^2 / (4\pi^2) \approx 0.025$~\AA$^2$, which corresponds to a typical order of magnitude of the MSD of atoms in a material at room temperature. For each displacement-averaged multislice simulation scenario we sampled 8192 different displacements of the carbon atom based on the specific prescription for that simulation. From these displacements one was chosen at random 81920 times to compute the corresponding exit wave function using the \texttt{DrProbe} software package \cite{barthel_dr_2018}. From these exit wave functions we extract the coherent and inelastic parts of the intensity, c.f.\ Sec.~\ref{sec:FRFPMS_in_WPOA}. The simulation box size is set to $4 \times 4 \times 0.3$~nm$^3$ and the grid on which multislice calculations were run is $512\times 512$~pixels. We assume the projection approximation for the scattering potential, which is parametrized by the elastic atomic scattering factor for carbon of Waasmeier and Kirfel \cite{waasmaier_new_1995} and the entire scattering action of the atom is then contained in a single slice of thickness 0.3~nm. The incident electron beam has a kinetic energy of 60~keV and we set the convergence semi-angle to 1~mrad, producing a near parallel beam. The beam is furthermore centered on the equilibrium position of the carbon atom.

\begin{figure}
    \centering
    \includegraphics[width=\linewidth]{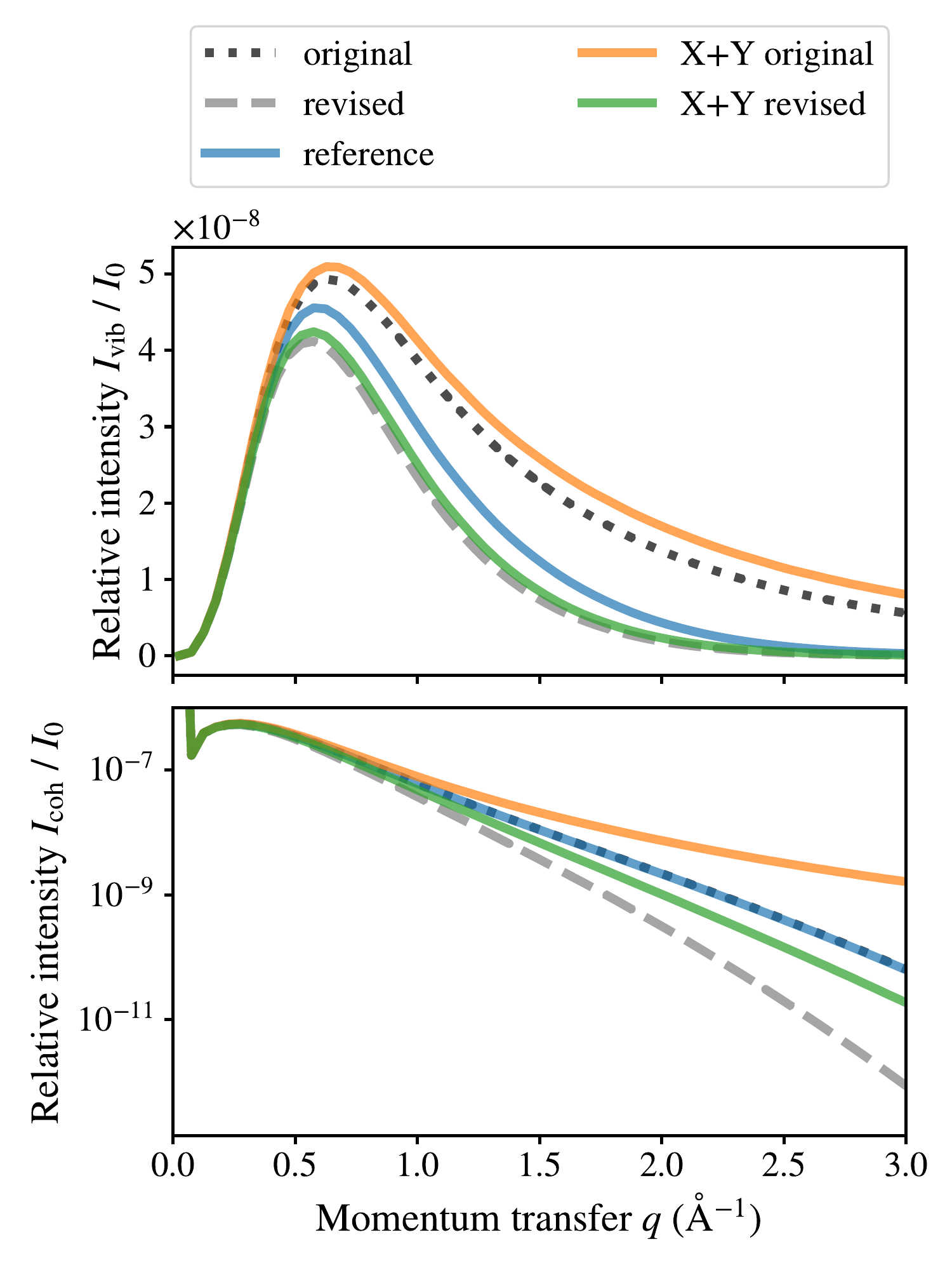}
    
    \caption{Overview of the numerical simulations for a value of the MSD $\langle \pmb{\tau}^2 \rangle_T \approx 0.025$~\AA$^2$: the top panel displays the azimuthally integrated, inelastically scattered relative intensity and the bottom panel the azimuthally integrated relative coherent intensity for the five considered calculations. These include the \emph{original} and the \emph{revised} FRFPMS method, labeled ``original'' and ``revised'' for the combined $x$- and $y$-displacements calculations above. The sum of the results of calculations considering separate $x$- and $y$-displacements are labelled by an additional ``X+Y'' for both editions of the FRFPMS method. The reference results labelled ``reference'' are obtained using Eqs.~\eqref{eq:ref_inelastic} and \eqref{eq:ref_elastic}.
    }
    \label{fig:Figures/radial_integrals_inela_ela_Biso0.010_dxy10_cmpref_base}
\end{figure}

We repeat the calculation once for the \emph{original} FRFPMS method and once for the \emph{revised} FRFPMS method. We reiterate here, that the difference between these versions is purely the inclusion of the total DWF according to Eq.~\eqref{eq:FRFPMS_correction_DWF}. The IQHO has two modes, which occur at the same frequency, so they are part of the same frequency bin and the \emph{original} FRFPMS method calculation is then equivalent to a standard QEP/FPMS calculation from a practical perspective. In addition to these calculations, we repeat the calculation also for a situation, in which the carbon atom is displaced along one of the dimensions of the IQHO for both the \emph{original} and the \emph{revised} FRFPMS method. While this calculation is artificial for the IQHO, it lets us consider the effect of separating the modes into separate frequency bins, as if they occurred at different frequencies. This can be thought of as the limiting case, when $\omega_x$ and $\omega_y$ are in different energy bins, while approaching the same value $\omega_x,\omega_y \rightarrow \omega$. From a different point of view, nothing prevents us to realize a frequency- and mode-resolved FPMS method.

In all calculations, we extract the total, coherent and inelastically scattered intensities through incoherent and coherent averages of the exit wave function (c.f. Eq.~\ref{eq:FRFPM_Iincoh_Icoh_Iinel}) and compare these results with formulas based on the first-order Born approximation in Eq.~\eqref{eq:single_inel_cross_section_aniso_qmharm_osci}. We then compare the inelastically scattered intensity with the reference expression
\begin{equation}
    \label{eq:ref_inelastic}
    \begin{aligned}
        I_{\mathrm{vib}}^{(\mathrm{ref})}(\mathbf{q}) 
        = {} & \frac{2\pi^2 \gamma^2 \hbar}{M L^2 k_0^2} \frac{[1 + 2\langle n_x \rangle_T]}{\omega} \; q_x^2 \; f_e^2(q)  e^{-2W(\mathbf{q})} \\
        {} + {} & \frac{2\pi^2 \gamma^2 \hbar}{M L^2 k_0^2} \frac{[1 + 2\langle n_y \rangle_T]}{\omega} \; q_y^2 \; f_e^2(q)  e^{-2W(\mathbf{q})} \\
        = {} & \frac{4 \pi^2 \gamma^2}{L^2 k_0^2} \, q^2 \left\langle \tau^{2} \right\rangle_T \, f_e^2(q) \, e^{-2W(q)},
    \end{aligned}
\end{equation}
where $\langle\tau^2 \rangle_T = \left\langle \tau_x^{2}\right\rangle_T = \left\langle \tau_y^{2}\right\rangle_T$ is the isotropic equivalent of the MSD of the IQHO (c.f. Eq.~\eqref{eq:FRFPMS_MSD_qm_factor2}). Furthermore we have added the contributions of energy-loss and energy-gain processes, and introduced the relativistic $\gamma$-factor to be consistent with our treatment of the FRFPMS method. We compare the coherent intensity with the reference expression
\begin{equation}
    \label{eq:ref_elastic}
    I_{\mathrm{coh}}^{(\mathrm{ref})}(\mathbf{q}) = \frac{1}{L^2} \delta(\mathbf{q}) + \frac{\gamma^2}{L^2 k_0^2} f_e^2(q) \, e^{-2W(q)},
\end{equation}
which follows from an analogous calculation to that which lead to equation~\eqref{eq:FRFPMS_AQHO_Icoh}, albeit using both modes of the IQHO instead of only the $\omega_x$-mode.

We display the comparison of the inelastically scattered intensities in our numerical evaluation in the top panel of Fig.~\ref{fig:Figures/radial_integrals_inela_ela_Biso0.010_dxy10_cmpref_base}. Generally the \emph{original} FRFPMS method gives the largest inelastically scattered intensity with zero intensity at zero momentum transfer, a pronounced maximum at intermediate momentum transfers and decreasing intensity at large momentum transfers. At small momentum transfers below about 0.5~\AA$^{-1}$, all calculations yield similar results, which can be understood in terms of the criterion in Eq.~\ref{eq:FRFPMS_AQHO_1st_order_requirement}, which has a value of $2W(\mathbf{q}) \approx 0.12$ at $q=0.5$~\AA$^{-1}$. At intermediate momentum transfers, we obtain a similar shape of the inelastically scattered intensity, but the maximum of the inelastic intensity is smaller and occurs at slightly smaller momentum transfers in the \emph{revised} FRFPMS compared with the \emph{original} FRFPMS. The results of the reference calculation are somewhat between the results of both editions of the FRFPMS method. Furthermore the exponential decrease at large momentum transfer differs between the calculations: the \emph{revised} FRFPMS method calculation exhibits a steeper decrease than the reference and the \emph{original} FRFPMS method. Separately considering the displacements along $x$ and $y$ in the calculations yields almost no change for the inelastically scattered intensities for both editions of the FRFPMS method. Overall, we deem the \emph{revised} FRFPMS a modest improvement in terms of the inelastically scattered intensity, since it matches the reference expectation better than the \emph{original} FRFPMS method.

In the bottom panel of Fig.~\ref{fig:Figures/radial_integrals_inela_ela_Biso0.010_dxy10_cmpref_base}, we display the results for the coherent intensity. As mentioned before, the coherent intensity in the \emph{original} FRFPMS method is equivalent to a QEP/FPMS calculation and it is therefore the correct coherent intensity for the considered situation, as the excellent agreement with the reference result shows. The coherent intensity in the \emph{revised} FRFPMS method on the other hand agrees well at small momentum transfers, but deviates from the other two calculations for momentum transfers larger than about 0.5~\AA$^{-1}$. Separately considering the displacements of both modes of the IQHO has a rather strong effect on the coherent intensity in the \emph{original} FRFPMS method, which overestimates elastic scattering towards large momentum transfers in this situation, while the coherent intensity in the \emph{revised} FRFPMS reproduces the coherent reference intensity much better.

Overall these results show, that the \emph{revised} FRFPMS method is, for the considered model, more successful in modelling the inelastic intensity, while the \emph{original} FRFPMS method is more accurate in modeling the elastic scattering for the IQHO. This would suggest that the \emph{revised} FRFPMS would only be applicable to very thin samples, where dynamical diffraction is negligible. So where is the promised \emph{improvement} by including the explicit DWF smearing? A hint at the answer to this question can be found in the calculations considering separate $x$ and $y$-displacements. Neglecting anisotropies, what we effectively achieve in these calculations is, that the total MSD $\langle \pmb{\tau}^2 \rangle_T$ is \emph{smaller} than it is in the calculations which consider displacements along both directions simultaneously. We have seen, that the \emph{revised} FRFPMS method performed better under these conditions than the \emph{original} FRFPMS method, both for modelling the inelastically scattered as well as the coherent intensities. Extrapolating this observation, we expect that as the displacements of atoms decrease in a given frequency bin, the \emph{revised} FRFPMS method will agree better and better with the reference calculation, effectively fulfilling the criterion of Eq.~\eqref{eq:FRFPMS_AQHO_1st_order_requirement} better.

\begin{figure}
    \centering
    \includegraphics[width=\linewidth]{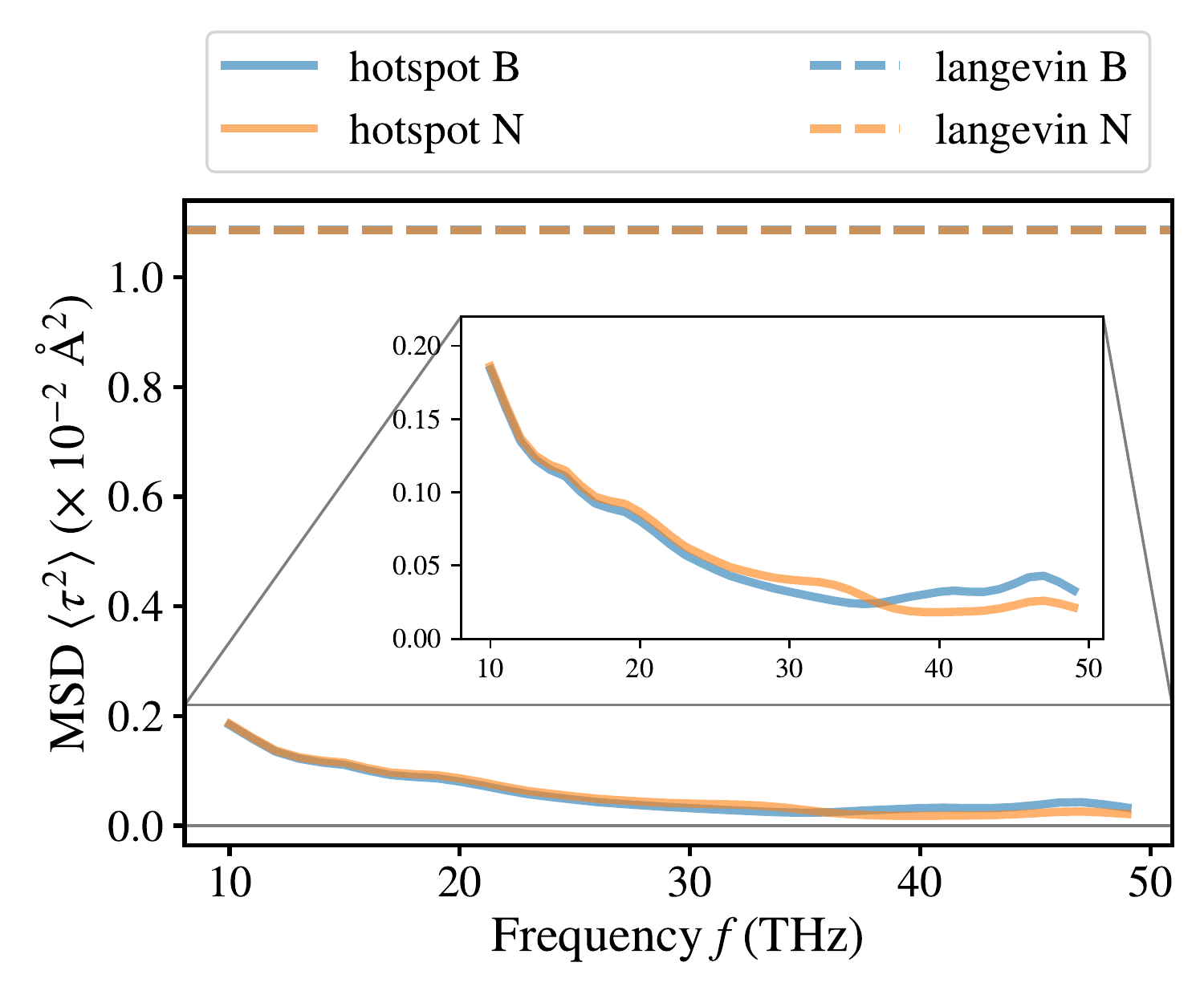}
    \caption{Comparison between the frequency-resolved MSD in $x$ and $y$-directions in hotspot thermostat MD simulations and the total MSD along $x$ and $y$-directions in a constant temperature MD simulation using a Langevin thermostat. Note that the MSD in the MD calculation using the Langevin thermostat has contributions from all modes (frequencies). The simulated material is a crystal of hBN, whose $c$-axis is aligned with the $z$-direction. More details about the underlying MD simulations can be found in Refs.~\cite{zeiger_frequency-resolved_2021,zeiger_simulations_2021}.}
    \label{fig:Figures/msd_comparison_xy}
\end{figure}

Herein, in the \emph{size} of the displacements in any real FRFPMS calculation, lies the answer to the question about where the \emph{improvement} of the \emph{revised} FRFPMS over the \emph{original} FRFPMS is. So far, we have considered a MSD, which is typical for a \emph{total} MSD of an atom at room temperature, which is the natural procedure if one is modelling the effect of all vibrational modes together at the same time. The AQHO is, however, a rather artificial model for the FRFPMS method, since the PDOS $g(\omega)$ of this model is basically a Dirac-$\delta$ at the frequency of the $x$ and $y$-modes, i.e.,
\begin{equation}
    g(\omega) \propto \delta(\omega-\omega_x) + \delta(\omega-\omega_y),
\end{equation}
where $\omega_x=\omega_y$ for the IQHO. The frequency-resolved thermostat in a FRFPMS calculation would then either excite the modes with their full MSD $\left\langle \pmb{\tau}^2 \right\rangle_T$, if the peak frequency of the thermostat is tuned to the frequency $\omega_x = \omega_y$, or keep them ``frozen'', if it is tuned away from the IQHO mode frequency. In other words, all possible vibrational modes contribute to the same frequency bin for the IQHO.

In any real material, the PDOS is, however, a broad function of frequency and every atom participates in vibrations within a (wider) range of frequencies. The frequency-resolved thermostat in the MD simulation allows only a subset of all modes to contribute to the displacements of any atom, while most frequencies are suppressed. This leads to reduced, \emph{effective} MSDs in every frequency bin. Figure~\ref{fig:Figures/msd_comparison_xy} illustrates this circumstance based on trajectory data originally simulated for hBN in connection with Refs.~\cite{zeiger_frequency-resolved_2021,zeiger_simulations_2021}. In Fig.~\ref{fig:Figures/msd_comparison_xy} the \emph{effective} MSD is much lower for any of the frequencies of the hotspot thermostat than the total MSD in a constant temperature MD simulation, which involves all frequencies and enforces the equipartition of energy between all available modes. 

Conveniently, when one adds the energy-loss and energy-gain cross-section, the inelastic cross-section is only a function of MSD, see Eq.~\eqref{eq:FRFPMS_Ivib_final} as well as Eq.~\eqref{eq:single_inel_cross_section_aniso_qmharm_osci_T0K} (for $T=0$~K) and Eq.~\eqref{eq:single_inel_cross_section_aniso_qmharm_osci} (for nonzero $T$, where we need to add energy-gain processes, see also Eq.~\eqref{eq:frfpms_energy_scaling_2n+1_limits}). This allows one to model a situation in which the displacements $\pmb{\tau}$ are \emph{reduced} compared to the total MSD $\left\langle \pmb{\tau}^2 \right\rangle_T$. In particular, we consider displacements which correspond to an \emph{effective} MSD of 16\% of the total MSD, i.e., $\langle \pmb{\tau}^2 \rangle_T \rightarrow 0.16 \cdot 0.025$~\AA$^2 = 0.4\cdot 10^{-2}$~\AA$^2$. This value of the MSD follows the largest value of the \emph{effective} MSDs in the frequency bins in Fig.~\ref{fig:Figures/msd_comparison_xy} rather well. We then calculate again the inelastically scattered and the coherent intensities within the frameworks of the \emph{original} and \emph{revised} FRFPMS methods for displacements along both directions as well as for separate displacements. The DWF-smearing is thereby kept at its nominal value, corresponding to the total MSD, and only the atomic displacements are reduced. We then compare the inelastically scattered intensity with the expression
\begin{equation}
    \label{eq:ref_inelastic_04}
    \begin{aligned}
        \tilde{I}_{\mathrm{vib}}^{(\mathrm{ref})}(\mathbf{q}) 
        = {} & 0.16 \times \frac{4 \pi^2 \gamma^2}{L^2 k_0^2} \, q^2  \left\langle \tau^{2} \right\rangle_T \, f_e^2(q) \, e^{-2W(q)} \\
        = {} & 0.16 \times I_{\mathrm{vib}}^{(\mathrm{ref})}(\mathbf{q}),
    \end{aligned}
\end{equation}
and Eq.~\eqref{eq:ref_elastic} for the coherent intensity as for full displacements.

\begin{figure}
    \centering
    \includegraphics[width=\linewidth]{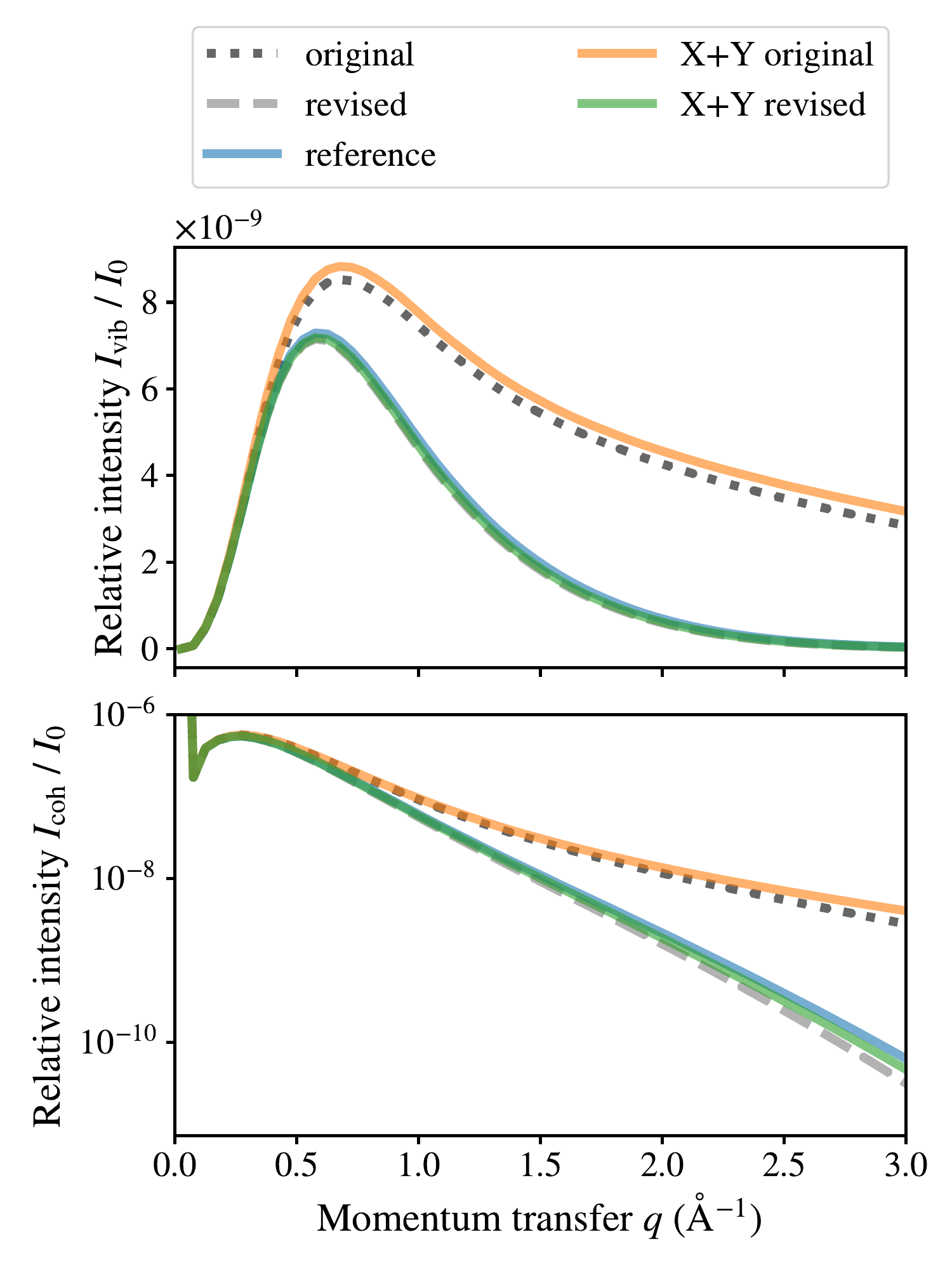}
    \caption{Comparison of inelastically scattered intensity (top panel) and coherent intensity for different approaches to phonon scattering similar to Fig.~\ref{fig:Figures/radial_integrals_inela_ela_Biso0.010_dxy10_cmpref_base}. Here the explicit displacements were scaled down, such that they correspond to 16\% of the MSD $\langle \pmb{\tau}^2 \rangle_T$ of the oscillator and the reference result is then given by Eqs.~\eqref{eq:ref_inelastic_04} for inelastic scattering and continues to be given by Eq.~\ref{eq:ref_elastic} for the coherent intensity.}
    \label{fig:Figures/radial_integrals_inela_ela_Biso0.010_dxy04_cmpref_base}
\end{figure}

The results for this \emph{reduced} displacement case are displayed in Fig.~\ref{fig:Figures/radial_integrals_inela_ela_Biso0.010_dxy04_cmpref_base}. The overall shape of the inelastically scattered intensity is similar to the large displacement case in Fig.~\ref{fig:Figures/radial_integrals_inela_ela_Biso0.010_dxy10_cmpref_base}, but the maximum intensity is about one order of magnitude lower. At large momentum transfers, the inelastic intensity is much larger in the \emph{original} FRFPMS method than for the \emph{revised} FRFPMS method and the reference calculation, which agree very well with each other. This suggests, that the \emph{original} FRFPMS method overestimates large angle inelastic scattering, whereas the \emph{revised} FRFPMS method is much closer to the reference based on single inelastic scattering theory. Separating the displacements along $x$ and $y$ does not change this picture for the \emph{original} or the \emph{revised} FRFPMS method.

The \emph{revised} FRFPMS method not only improves the description of inelastic scattering, but also the coherent intensity, shown in the bottom panel of Fig.~\ref{fig:Figures/radial_integrals_inela_ela_Biso0.010_dxy04_cmpref_base}, agrees very well with the coherent reference intensity. Only at large momentum transfers, beyond about 1.5~\AA$^{-1}$, do we observe a small difference between the reference result and the \emph{revised} FRFPMS method. The \emph{original} FRFPMS method overestimates the coherent intensity quite severely at momentum transfers larger than about 0.75~\AA$^{-1}$. Again, separating the displacements along $x$ and $y$ has no appreciable effect for both editions of the FRFPMS method.

Furthermore the results in Fig.~\ref{fig:Figures/radial_integrals_inela_ela_Biso0.010_dxy04_cmpref_base} show that the accuracy of the inelastically scattered intensity for the \emph{revised} FRFPMS method is in principle not limited by the criterion of Eq.~\eqref{eq:FRFPMS_AQHO_1st_order_requirement}. The inclusion of the DWF leads to a dramatic increase of the accuracy of the inelastic intensity as a function of momentum transfer and also decreases large momentum transfer elastic scattering in line with the reference calculation. Comparing Figs.~\ref{fig:Figures/radial_integrals_inela_ela_Biso0.010_dxy10_cmpref_base} and \ref{fig:Figures/radial_integrals_inela_ela_Biso0.010_dxy04_cmpref_base} suggests, that the relative agreement between the \emph{revised} FRFPMS method and the reference result becomes purely a function of the \emph{size} of the atomic displacements. Indeed, we have verified (not shown here), that we obtain a near perfect agreement for the coherent intensity at all considered momentum transfers, if the displacements are further reduced, such that they are consistent with 1\% of the MSD of the IQHO considered here, i.e. $\langle \pmb{\tau}^2 \rangle_T = 2\langle \tau^2 \rangle_T = 2.5 \cdot 10^{-4}$~\AA$^2$. Also the inelastically scattered intensity continues to be basically indistinguishable from the reference calculation for such further reduced displacements. Overall this suggests, that the number of vibrational states within a frequency-bin is the main criterion determining the accuracy for the \emph{improved} FRFPMS calculation. Increasing the number of bins, such that fewer modes contribute to each individual bin leads to smaller displacements within each bin and thus to more accurate results. This suggests an alternative avenue of improving the precision of single-phonon inelastic scattering in \emph{revised} FRFPMS calculations, namely by scaling down the displacements followed by an a posteriori scaling up of the calculated inelastic intensities. We leave this as an option for future studies.

\section{Conclusions and Outlook}
\label{sec:conclusion_outlook}

We have shown on the example of the AQHO, that the first-order Born approximation, the QEP model, and the FRFPMS method result in the same cross section for the single inelastically scattered intensity as a function of momentum transfer. These considerations led furthermore to the result that the \emph{original} FRFPMS method underestimates the amount of DWF smearing of the potential, since only a subset of all possible modes is present in any frequency bin. This led to the proposal of a \emph{revised} prescription for the FRFPMS method by including an \emph{explicit} DWF smearing in the displaced potential, in which we break the strict classical separation between two types of electron scattering calculations, which take thermal vibrations into account exclusively via displacement-averaging or via DWF-smeared potentials, respectively. We reiterate at this point, that the intention with the FRFPMS method is different from these other types of electron scattering calculations: we specifically want to reproduce the \emph{single} inelastically scattered intensity and the full elastically scattered intensity.

We have shown that numerical simulations presented in Sec.~\ref{sec:numerical_sims} using such a \emph{revised} FRFPMS method are in good agreement with reference calculations of single inelastic phonon scattering and elastic scattering. To be explicit the \emph{revised} FRFPMS method has the following advantages over the \emph{original} FRFPMS method: in the range of considered momentum transfers $0 <q <3$~\AA$^{-1}$ it predicts the amount of \emph{single} inelastic scattering in the considered frequency bin much better and also improves the description of elastic scattering due to the inclusion of an \emph{explicit} DWF smearing of the potential. The numerical simulations also illustrate that the only requirement for the \emph{revised} FRFPMS method to be accurate is that the displacements of any given atom do not get too large in any frequency bin, see Eq.~\eqref{eq:FRFPMS_AQHO_1st_order_requirement}. %

An issue, which we have not considered in this work is absorption. Any FRFPMS method will inherently account incorrectly for the total amount of intensity absorbed from the elastic channel. The absorbed intensity will always be smaller than in a full QEP/FPMS calculations, since only a subset of all modes contribute to the inelastic intensity in an energy bin. Ideally we would like to describe the same elastic intensity in every frequency bin, which could be achieved by including an imaginary (absorptive) part in the scattering potential \cite{weickenmeier_computation_1991,martin_model_2009}. We leave this question and any details connected to it open for future investigations.

\begin{acknowledgments}
We thank Axel Lubk for helpful discussions and critical feedback on the manuscript. Furthermore we acknowledge the support of the Swedish Research Council, Olle Engkvist’s Foundation, Carl Trygger's Foundation, Knut and Alice Wallenberg Foundation, and eSSENCE for financial support. Simulations were enabled by resources provided by the Swedish National Infrastructure for Computing (SNIC) at NSC Centre partially funded by the Swedish Research Council through Grant Agreement No. 2018-05973.
\end{acknowledgments}

\appendix

\section{Properties of the QHO, its wave functions and Hermite polynomials}
\label{app:properties_qho_wf}

Consider a quantum mechanical particle of mass $M$ in a quadratic potential, i.e., the Hamiltonian
\begin{equation}
    H = \frac{p^2}{2M} + \frac{M\omega^2 x^2}{2},
\end{equation}
where $p=-i\hbar\partial/\partial x$ and $x$ are the momentum and position operator, respectively, $\hbar=h/2\pi$ is the reduced Planck constant, and $\omega^2$ is the frequency of the oscillator. The Schrödinger equation of the system then reads
\begin{equation}
    \label{app_eq:SEQ_qmharm_osci}
    -\frac{\hbar^2}{2M} \frac{\partial^2}{\partial x^2} a(x) + \left[ \frac{M\omega^2 x^2}{2} - E \right] a(x) = 0
\end{equation}
where $a(x)$ is the particle's wave function. It turns out, that Eq.~\eqref{app_eq:SEQ_qmharm_osci} has solutions
\begin{equation} \label{app_eq:a_n_qmharm_osci}
    a_n(x) 
    = \frac{1}{\sqrt{2^n\,n!}}  \left(\frac{M\omega}{\pi \hbar}\right)^{\frac{1}{4}} \exp\left[- \frac{M\omega x^2}{2 \hbar}\right] H_n\left(\sqrt{\frac{M\omega}{\hbar}} x \right)
\end{equation}
for $n = 0,1,2,\ldots$, where
\begin{equation}
    \label{app_eq:Hermite_polys_definition}
    H_n(x)=(-1)^n ~ e^{x^2}\frac{d^n}{dx^n}\left( e^{-x^2} \right)
\end{equation}
are the ``Physicist's Hermite polynomials''. The Hermite polynomials satisfy the following orthogonality relation
\begin{equation}
    \label{app_eq:orthogonality_hermite_polynomials}
    \int_{-\infty}^\infty H_m(x) H_n(x)\, e^{-x^2} \,dx = \sqrt{\pi}\, 2^n\, n!\, \delta_{nm}
\end{equation}
and the recursion relation
\begin{equation}
    \label{app_eq:H_n_recursion}
    H_{n+1} (x) = 2x H_n(x) - 2n H_{n-1}(x)
\end{equation}
with $H_0(x) = 1$ and $H_1(x) = 2x$. Furthermore, we also have the following result \cite[p. 194 eq. 10.13]{erdelyi_higher_1953-2}
\begin{equation}
    \label{app_eq:sum_zn_n!_Hnx_Hny}
    \sum_{n=0}^\infty \left(\frac{z}{2}\right)^n \frac{H_n(x) H_n(y)}{n!}  = \frac{1}{\sqrt{1 - z^2}} e^{\frac{2z}{1 + z}xy - \frac{z^2}{1 - z^2}(x - y)^2}
\end{equation}
for $z<1$. The following integral involving Hermite polynomials will prove useful in later appendices:
\begin{equation} \label{app_eq:int_e-x2_Hn_Hm_eibetax}
    \begin{aligned}
        \int_{-\infty}^\infty \frac{e^{-x^2} H_n(x) H_m(x)}{\sqrt{2^n n! 2^m m! \pi}} {} & {} e^{i\beta x}dx = {} \\
        =  \sqrt{\frac{2^m m!}{2^n n!}} (i\beta)^{n-m} {} & {} L_m^{n-m}\left(\frac{\beta^2}{2}\right) e^{-\frac{\beta^2}{4}},
    \end{aligned}
\end{equation}
where $L_m^{n-m}\left(x\right)$ are the generalized Laguerre polynomials, for which the following relation holds for integer $m,n$:
\begin{equation}
    \frac{(-x)^m}{m!}L_n^{m-n}(x) = \frac{(-x)^n}{n!}L_m^{n-m}(x).
\end{equation}
Another useful result concerns the sum of generalized Laguerre polynomials \cite{martin_model_2009}:
\begin{equation} \label{app_eq:lagident}
    \begin{aligned}
        \sum_{n=0}^\infty \frac{n!}{(m+n)!} {} & {} z^n L_n^m(x) L_n^m(y) = {} \\
        = {} & {} \frac{e^{-(x+y)\frac{z}{1-z}}}{1-z} \frac{I_m\left(2\sqrt{xy} \frac{\sqrt{z}}{1-z} \right)}{(xyz)^{m/2}},
    \end{aligned}
\end{equation}
where $m$ is a positive integer and $I_m$ are the modified Bessel functions of the first kind.

The MSD of the harmonic oscillator reads
\begin{equation}
    \label{app_eq:properties_QHO_MSD}
    \begin{aligned}
        \left\langle x^2 \right\rangle 
        = {} & {} \frac{e^{\beta \hbar \omega} - 1}{e^{\beta \hbar \omega}} \sum_{n=0}^{\infty} e^{-\beta \hbar \omega n} \int dx \; x^2 a_n^*(x) a_n(x) \\
        = {} & {} \left(\frac{\hbar}{2M\omega}\right) \coth\left(\frac{\beta \hbar \omega}{2}\right),
    \end{aligned}
\end{equation}
at temperature $T = 1/(k_{\mathrm{B}}\beta)$. Part of the expression in front of the sum is coming from the statistical sum $Z = \sum_{n=0}^\infty e^{-\beta \hbar \omega (n+\frac{1}{2})}$.

\section{Phonons in harmonic approximation}
\label{app:phonons_in_harm_approx}

The displacement of atom $j$ in unit cell $l$ in harmonic approximation and with Born-von Karman periodic boundary conditions reads \cite{maradudin_theory_1963}
\begin{eqnarray}
    \label{app_eq:harmonic_approx_atomic_displacement}
    \mathbf{u}_{jl}(t) & = & \sqrt{ \frac{\hbar}{2 \mathcal{N}_{\mathrm{uc}} m_{j}}} \sum_{\mathbf{q} \nu} \frac{1}{\sqrt{\omega (\mathbf{q}, \nu)}} e^{i\mathbf{q}\pmb{\cdot} \mathbf{R}_{jl}} \; \pmb{\epsilon}_{j}(\mathbf{q}, \nu) \nonumber \\
    & \times & \left[ \hat{a} (\mathbf{q}, \nu) e^{-i \omega (\mathbf{q}, \nu) t} + \hat{a}^{\dagger} (-\mathbf{q}, \nu) e^{i \omega (\mathbf{q}, \nu) t} \right],
\end{eqnarray}
where $\hat{a} (\mathbf{q}, \nu)$ and $\hat{a} (\mathbf{q}, \nu)^{\dagger}$ are the phonon creation and annihilation operators for mode $(\mathbf{q}, \nu)$, respectively, and the normalization is given by the number of unit cells $\mathcal{N}_{\mathrm{uc}}$. The phonon polarization vectors satisfy
\begin{equation}
    \label{app_eq:phonons_in_harm_approx_phonon_polarization_conjugation_property}
    \pmb{\epsilon}_j(-\mathbf{q}\nu) = \pmb{\epsilon}_j^*(\mathbf{q}\nu).
\end{equation}
The MSD at temperature $T=(k_{\mathrm{B}} \beta)^{-1}$ then reads
\begin{equation}
\label{app_eq:harmonic_approx_atomic_msd}
\begin{aligned}
    \left\langle \left| \mathbf{u}_{jl}(t) \right|^2 \right\rangle_T 
    = {} & \frac{\hbar}{2 \mathcal{N}_{\mathrm{uc}} m_{j}} \sum_{\mathbf{q} \nu} \frac{1 + 2 \, \langle n(\mathbf{q}, \nu) \rangle_T}{\omega (\mathbf{q}, \nu)} \left| \pmb{\epsilon}_{j}(\mathbf{q}, \nu) \right|^2 \\ 
    = {} & \frac{\hbar}{2 \mathcal{N}_{\mathrm{uc}} m_{j}} \sum_{\mathbf{q} \nu} \frac{\coth\left(\beta \hbar\omega (\mathbf{q}, \nu)\right)}{\omega (\mathbf{q}, \nu)} \left| \pmb{\epsilon}_{j}(\mathbf{q}, \nu) \right|^2,
\end{aligned}
\end{equation}
where $\langle n(\mathbf{q}, \nu) \rangle_T = 1/(e^{\beta\hbar\omega (\mathbf{q}, \nu)}-1)$ is the average occupation number of phonon mode $(\mathbf{q}, \nu)$. One can furthermore define a projected MSD along some unit vector $\mathbf{\hat{n}}$
\begin{equation}
    \label{app_eq:harmonic_approx_atomic_msd_projected}
    \begin{aligned}
        \left\langle \left| \mathbf{\hat{n}}\pmb{\cdot} \mathbf{u}_{jl}(t) \right|^2 \right \rangle_T {} & {} =  \\
        = \frac{\hbar}{2 \mathcal{N}_{\mathrm{uc}} m_{j}} {} & {} \sum_{\mathbf{q} \nu} \frac{1 + 2 \, \langle n(\mathbf{q}, \nu) \rangle_T}{\omega (\mathbf{q}, \nu)} \left| \mathbf{\hat{n}} \pmb{\cdot} \pmb{\epsilon}_{j}(\mathbf{q}, \nu) \right|^2.
    \end{aligned}
\end{equation}
For isotropic oscillators, one has
\begin{equation}
    \label{app_eq:relation_MSD_isotropicequiv_MSD}
    \left\langle \left| \mathbf{u}_{jl}(t) \right|^2 \right\rangle_T = 3 \left\langle \left| \mathbf{\hat{n}}\pmb{\cdot}\mathbf{u}_{jl}(t) \right|^2 \right\rangle_T
\end{equation}
for any unit vector $\mathbf{\hat{n}}$. We use this property in the following appendices.

\section{Debye-Waller factor}
\label{app:DWF}

In a solid, atoms undergo thermal motion at finite temperature $T$, i.e., the position of the $i$-th atom is a function of time $\mathbf{R}_i(t)$. In crystals, such motion happens around the equilibrium position of the atom defined as the time averaged position of the atom $\left\langle \mathbf{R}_i(t) \right\rangle$. Thus we can write the instantaneous position  $\mathbf{R}_i(t)$ as
\begin{equation}
    \mathbf{R}_i(t) = \left\langle \mathbf{R}_i(t) \right\rangle + \mathbf{U}_i(t),
\end{equation}
where $\mathbf{U}_i(t)$ is the instantaneous displacement from equilibrium. In scattering experiments, the time of collecting enough counts for a reliable measurement of the scattering cross section is typically much longer than the time scale on which $\mathbf{U}_i(t)$ fluctuates. Thus for many applications in elastic scattering, it is sufficient to consider scattering only on the time averaged potential and the DWF arises precisely from this time averaging of the displaced atomic scattering potential. Equivalently one can also consider the thermodynamic average $\langle \ldots \rangle_T$, i.e.,
\begin{equation}
    \label{app_eq:DWF_origin}
    \begin{aligned}
        \left\langle V(\mathbf{Q}, \mathbf{U}) \right\rangle_T 
        = & {} \left\langle V(\mathbf{Q}) \; e^{2\pi i \mathbf{Q}\pmb{\cdot}\mathbf{U}} \right\rangle_T \nonumber\\
        = & {} V(\mathbf{Q}) \; e^{- 2 \pi^2 \left\langle (\mathbf{Q}\pmb{\cdot}\mathbf{U})^{2}\right\rangle_T},
    \end{aligned}
\end{equation}
where $V(\mathbf{Q})$ is the non-displaced scattering potential. The proof that the right hand side of Eq.~\eqref{app_eq:DWF_origin} is equal to the left hand side in the harmonic approximation can for example be found in chapter 7, section 2 of Ref.~\cite{maradudin_theory_1963}, where it is shown that
\begin{equation}
    \label{app_eq:<expiqu>}
    \left\langle e^{2\pi i \mathbf{Q}\pmb{\cdot}\mathbf{U}} \right\rangle_T = e^{- 2 \pi^2 \left\langle (\mathbf{Q}\pmb{\cdot}\mathbf{U})^{2}\right\rangle_T}.
\end{equation}
We then define
\begin{equation}
    \label{app_eq:DWF_definition}
    e^{-2 \pi^2 \left\langle (\mathbf{Q}\pmb{\cdot}\mathbf{U})^{2}\right\rangle_T} =: e^{- W(T,\mathbf{Q})},
\end{equation}
called the DWF, a function depending on temperature $T$ and momentum transfer $\mathbf{Q}$. In this paper we simplify the notation by writing $W(\mathbf{Q})$, when we mean $W(T,\mathbf{Q})$, and use an explicit index ``$0$'', if we mean $W(0,\mathbf{Q})$, i.e., $W_{0}(\mathbf{Q}) = W(0,\mathbf{Q})$. We can get a further insight into the meaning of the DWF when we consider 
\begin{equation}
    W(\mathbf{Q}) = 2 \pi^2 \left\langle (\mathbf{Q}\pmb{\cdot}\mathbf{U})^{2}\right\rangle_T = 2 \pi^2 Q^2 \left\langle \left(\hat{\mathbf{n}}_{\mathbf{Q}} \pmb{\cdot} \mathbf{U}\right)^{2}\right\rangle_T,
\end{equation}
where $\hat{\mathbf{n}}_{\mathbf{Q}} = \mathbf{Q}/Q$ is a unit vector in the direction of $\mathbf{Q}$. In other words, the function $W(\mathbf{Q})$ is the MSD at temperature $T$ projected along the direction of the momentum transfer (c.f. Eq.~\eqref{app_eq:harmonic_approx_atomic_msd_projected}) and multiplied by a factor $2\pi^2 Q^2$. For isotropic displacements, it follows, that
\begin{equation}
    2 \pi^2 Q^2 \left\langle (\mathbf{\hat{n}_{\mathbf{Q}}} \pmb{\cdot} \mathbf{U})^{2}\right\rangle_T = \frac{2 \pi^2}{3} Q^2 \left\langle \mathbf{U}^2 \right\rangle_T =: W(Q),
\end{equation}
where we have used Eq.~\eqref{app_eq:relation_MSD_isotropicequiv_MSD} and defined the isotropic function $W(Q)$. For anisotropic displacements, but with no correlation between vibrations along $x$, $y$, and $z$, i.e. $\left\langle u_i u_j\right\rangle = \delta_{ij} \left\langle u_{ii}^2\right\rangle$ with $i,j=x,y,z$, we find
\begin{equation}
\label{app_eq:DWF_anisotropic_no_correlation}
\begin{aligned}
    W(\mathbf{Q}) 
    {} & = 2\pi^2 \left\langle (\mathbf{Q}\pmb{\cdot}\mathbf{U})^{2}\right\rangle_T \\
    {} & =  2\pi^2 \left[ Q_x^2 \left\langle u_x^{2}\right\rangle_T + Q_y^2 \left\langle u_y^{2}\right\rangle_T + Q_z^2 \left\langle u_z^{2}\right\rangle_T \right] \\
    {} & =: W_x(Q_x) + W_y(Q_y) + W_z(Q_z),
\end{aligned}
\end{equation}
where $W_x(Q_x)$, $W_y(Q_y)$, and $W_z(Q_z)$ are associated with motion only along $x$, $y$, and $z$, respectively.

Similarly in 2D, the DWF becomes
\begin{equation}
    e^{-2 \pi^2 \left\langle (\mathbf{q}\pmb{\cdot}\mathbf{u})^{2}\right\rangle_T} =: e^{- W(T,\mathbf{q})},
\end{equation}
and for anisotropic, but uncorrelated displacements, we find
\begin{equation}
    W(q) = W_x(q) + W_y(q) = 2\pi^2 \left[ q_x^2 \left\langle u_x^{2}\right\rangle_T + q_y^2 \left\langle u_y^{2}\right\rangle_T \right].
\end{equation}
For convenience we define the following functions
\begin{subequations}
    \label{app_eq:DWF_2D_W0}
    \begin{align}
        W_{0x}(q_x) 
        = {} & {} 2 \pi^2 q_x^2 \left\langle u_{x}^2 \right\rangle_{0} 
        = q_x^2 \pi^2 \frac{\hbar}{\omega_x M}, \\
        W_{0y}(q_y)
        = {} & {} 2 \pi^2 q_y^2 \left\langle u_{y}^2\right\rangle_{0}
        = q_y^2 \pi^2 \frac{\hbar}{\omega_y M}
    \end{align}
\end{subequations}
associated with zero-point vibrations along $x$ and $y$, respectively.

\section{Transition potential approach by Forbes et al.}
\label{app:transition_potentials_Forbes}

We consider here the theory of phonon EELS as used by Forbes et al. \cite{forbes_modeling_2016} in order to compare it to Eq.~\eqref{eq:cross_section_finiteT_single_inelastic_scattering}. The theory considers only phonon excitation from the ground state $\mathbf{0}$ \cite{martin_model_2009} and the total cross section for inelastic scattering is expressed as (c.f. equation (4) of Ref. \cite{forbes_modeling_2016}),
\begin{eqnarray}
    \label{app_eq:forbes_et_al_sigma_transition_potential_reciprocal_space}
    \sigma & = & \frac{m^2}{4\pi^2\hbar^4} \sum_{\mathbf{n}} \frac{k_{\mathbf{n}}}{k_0} \int_{k'} \int_D \left|\int \psi_0(\mathbf{k}'-\mathbf{q}) H_{\mathbf{n}\mathbf{0}}(\mathbf{q}) \; \mathrm{d} \mathbf{q} \right|^2 \nonumber \\
    & \times & \delta(k_{\mathbf{n}} - k') \; \mathrm{d} \Omega_{k'} \mathrm{d}k'
\end{eqnarray}
where 
\begin{eqnarray} \label{app_eq:forbes_et_al_transition_potential_Hn0}
    H_{\mathbf{n}\mathbf{0}} (\mathbf{q}) 
    & = & \frac{2\pi^2\hbar^2}{m} \sum_{\kappa} f_{e}^{\kappa} (\mathbf{q}) ~ e^{- i \mathbf{q}\pmb{\cdot} \mathbf{R}_{\kappa}} \nonumber \\
    & \times & \prod_{j} \frac{\left[-i \pi \sqrt{2 M_{\kappa j}} \mathbf{q}\pmb{\cdot}\pmb{\epsilon}^{\kappa}_j\right]^{n_j}}{\sqrt{n_j!}} e^{-\pi^2 M_{\kappa j} (\mathbf{q}\pmb{\cdot}\pmb{\epsilon}_j^{\kappa})^2}
\end{eqnarray}
is the transition matrix element for the excitation $\mathbf{0}\rightarrow\mathbf{n} = (n_1, \ldots, n_j, \ldots)$ of the crystal and $M_{\kappa j} = \hbar/(m_{\kappa}\omega_j)$ and $\pmb{\epsilon}_j^{\kappa}$ is the phonon polarization vector of atom $\kappa$ in mode $j$.

Equation~\eqref{app_eq:forbes_et_al_sigma_transition_potential_reciprocal_space} describes the inelastic scattering cross section as the sum of incoherent contributions of transition potentials of the form of Eq.~\eqref{app_eq:forbes_et_al_transition_potential_Hn0} for each transition. The factor $k_{\mathbf{n}}/k_0$ is a signature of inelastic scattering and is connected to the fact that the incoming and outgoing probability currents flow with a different ``speed''. It should be noted, that Eq.~\eqref{app_eq:forbes_et_al_transition_potential_Hn0} is valid not only for single inelastic scattering, but explicitly allows for multi-phonon scattering from the ground state $\mathbf{0}$.

Furthermore, the transition potential approach as presented here is strictly speaking a $T=0$~K theory, since the crystal is assumed to be in the phonon ground state $\mathbf{0}$, as Eq.~(E3) in Ref.~\cite{martin_model_2009} signifies. There a generalized Laguerre function arises for any other initial state than the phonon ground state $\mathbf{0}$, which would in turn enter a generalization of Eq.~\eqref{app_eq:forbes_et_al_transition_potential_Hn0}. At finite temperature $T$ one would furthermore need to take a thermal average of the cross section over initial states, similar to what is calculated in Ref.~\cite{martin_model_2009} for the total inelastic absorption potential.

We turn our attention now to single phonon excitation from the ground state $\mathbf{n}_{j'} = (0,\ldots, 1, \ldots, 0)^T$, i.e. only one $n_{j'}=1$. We find from Eq.~\eqref{app_eq:forbes_et_al_transition_potential_Hn0}
\begin{eqnarray}
    H_{\mathbf{n}_{j'}\mathbf{0}} (\mathbf{q}) 
    & = & -\frac{h^2}{2m} \sum_{\kappa} e^{- i \mathbf{q}\pmb{\cdot} \mathbf{R}_{\kappa}} ~ f_{e}^{\kappa} (\mathbf{q}) ~ \sqrt{2M_{\kappa j'}} \; i \mathbf{q} \pmb{\cdot} \pmb{\epsilon}^{\kappa}_{j'} \nonumber \\
    & \times & \prod_{j} \exp\left[-\pi^2 M_{\kappa j'} (\mathbf{q}\pmb{\cdot}\pmb{\epsilon}_j^{\kappa})^2\right].
\end{eqnarray}
We can convert the product over phonon modes into a sum over phonon modes in the exponential and identify the DWF at $T=0$~K (c.f. Eqs.~\eqref{app_eq:harmonic_approx_atomic_msd} and \eqref{app_eq:DWF_definition})
\begin{equation}
    \label{eq:forbes_et_al_MSD_QM_transition_pot}
    \pi^2 \frac{\hbar}{m_{\kappa}} \sum_{j} \frac{1}{\omega_j} \left(\mathbf{q} \pmb{\cdot} \pmb{\epsilon}_j^{\kappa} \right)^2 = 2 \pi^2 \left\langle|\mathbf{q} \pmb{\cdot}\mathbf{u}_{\kappa}|^2\right\rangle_{0} = W_{0}(\mathbf{q}),
\end{equation}
where $ \left\langle|\mathbf{q} \pmb{\cdot}\mathbf{u}_{\kappa}|^2\right\rangle_{0}$ is the MSD at $T=0$~K. Thus we find for the transition potential
\begin{equation}
    \label{app_eq:forbes_et_al_Single_excitation_transition_potential_MSD}
    \begin{aligned}
        H_{\mathbf{n}_{j'}\mathbf{0}} (\mathbf{q}) 
        = {} & -\frac{2\pi^2\hbar^2}{m} \sum_{\kappa} \sqrt{2M_{\kappa j'}} \, e^{- i \mathbf{q}\pmb{\cdot} \mathbf{R}_{\kappa}}  \; i\mathbf{q}\pmb{\cdot}\pmb{\epsilon}^{\kappa}_{j'} \\
        {} & \phantom{-\frac{2\pi^2\hbar^2}{m} \sum_{\kappa}} \times f_{e}^{\kappa} (\mathbf{q}) e^{- W_{0}(\mathbf{q})}.
    \end{aligned}
\end{equation}
For a plane incident wave the inelastic scattering cross section can, according to Eq.~\eqref{app_eq:forbes_et_al_sigma_transition_potential_reciprocal_space}, been written as
\begin{equation}
\label{app_eq:forbes_et_al_Single_excitation_cross_section}
\begin{aligned}
    \frac{\mathrm{d}\sigma}{\mathrm{d}\Omega \mathrm{d}\omega} (\mathbf{q},\omega) = & 2\hbar \pi^2 \sum_{j'} \frac{k_{j'}}{k_0} \frac{1}{\omega_{j'}} \delta(\omega_{j'} - \omega) \times \\ 
    {} \times & \left| \sum_{\kappa} \frac{1}{\sqrt{m_{\kappa}}} \, e^{- i \mathbf{q}\pmb{\cdot} \mathbf{R}_{\kappa}} f_{e}^{\kappa} (\mathbf{q}) \; e^{- W_{0}(\mathbf{q})} \; \mathbf{q}\pmb{\cdot}\pmb{\epsilon}^{\kappa}_{j'} \right|^2,
\end{aligned}
\end{equation}
where $k_{j'}$ is the wave vector of the scattered inelastic wave. Eq.~\eqref{app_eq:forbes_et_al_Single_excitation_cross_section} exhibits the same functional dependencies as Eq.~\eqref{eq:cross_section_finiteT_single_inelastic_scattering} at temperature $T\rightarrow 0$, for which $\langle n(\mathbf{q}, \nu) \rangle_T\rightarrow 0$ and $W(\mathbf{q})\rightarrow W_{0}(\mathbf{q})$. Forbes et al. mention in Ref.~\cite{forbes_modeling_2016}, that the transition potential outlined here can be extended to non-zero temperatures by the reverse procedure, i.e. $W_{0}(\mathbf{q})\rightarrow W(\mathbf{q})$ and inserting $\coth(\beta\hbar\omega_{j'}) = 2\langle n_{j'} \rangle_T + 1$ before the Dirac-$\delta$ into Eq.~\eqref{app_eq:forbes_et_al_Single_excitation_cross_section}. A comparison with Eq.~\eqref{eq:cross_section_finiteT_single_inelastic_scattering} reveals, that such procedure would sum the energy-loss and -gain probabilities, but arrives otherwise at a compatible functional form.

As a last consideration, we note the standard result
\begin{equation}
\label{app_eq:fourier_transform_gradient}
    \mathcal{FT}\left[ \pmb{\nabla} f(\mathbf{r})\right] = 2\pi i \; \mathbf{q}\pmb{\cdot} \mathcal{FT}\left[f(\mathbf{r})\right],
\end{equation}
i.e., the Fourier transform $\mathcal{FT}[\ldots]$ of the gradient of a function is the product of $i\mathbf{q}$ and the Fourier transform of said function. If we Fourier transform Eq.~\eqref{app_eq:forbes_et_al_Single_excitation_transition_potential_MSD}, we can use this relationship in order to obtain the real-space interaction matrix element
\begin{align}
    H_{\mathbf{n}_{j'}\mathbf{0}} (\mathbf{r}) 
    = {} & \mathcal{FT}^{-1}\left[H_{\mathbf{n}_{j'}\mathbf{0}} (\mathbf{q}) \right] \\
    = {} & \sum_{\kappa} \sqrt{2M_{\kappa j'}} \; \pmb{\epsilon}^{\kappa}_{j'} \pmb{\cdot} \bm{\nabla}_{\mathbf{r}} \langle V(\mathbf{r})\rangle_{\mathbf{0}},
    \label{eq:Single_excitation_transition_potential_isotropic_MSD_structure_real_space}
\end{align}
which is a sum over gradients of the thermally averaged potential at $T=0$~K
\begin{equation}
     \langle V^{\kappa}(\mathbf{r})\rangle_{0} = -\frac{2\pi\hbar^2}{m} \; \mathcal{FT}\left[ f^{\kappa}(\mathbf{q}) e^{- i 2\pi \mathbf{q}\pmb{\cdot} \mathbf{R}_{\kappa}} e^{- W_{0}(\mathbf{q})} \right].
\end{equation}
A similar form of the inelastic potential was used by Dwyer in Eq.~(5) of Ref.~\cite{dwyerProspectsSpatialResolution2017}.

\begin{widetext}
\section{Solution of the integrals over AQHO wave functions in the QEP model}
\label{app:QEP_derivations_AQHO}

We consider here the second term of Eq.~\eqref{eq:QEP_AQHO_psi_nn+1_integrals_an_an+1_eiqtau}, i.e.,
\begin{equation}
    \label{app_eq:QEP_AQHO_psi_nn+1_integrals_an_an+1_eiqtau_2ndterm}
    \psi_{n_x, n_x+1}(\mathbf{q}) = \frac{1}{L} i\frac{\gamma}{k_0} f_e(q) \int a_{n_x+1}^*(\tau_x) a_{n_x}(\tau_x) \, e^{2\pi i q_x \tau_x} d\tau_x \, \int a_{n_y}^*(\tau_y) a_{n_y}(\tau_y) \, e^{2\pi i q_y \tau_y} d\tau_y.
\end{equation}
The wave functions are wave functions of the quantum harmonic oscillator, c.f. Eq.~\eqref{app_eq:a_n_qmharm_osci} and the first integral can thus be written in terms of Hermite polynomials, i.e.,
    \begin{equation}
    \begin{aligned}
        \label{app_eq:QEP_AQHO_integral_an+1n}
        \int a_{n_x+1}^*(\tau_x) a_{n_x}(\tau_x) \, e^{2\pi i q_x \tau_x} d\tau_x
        = {} & {} \frac{1}{\sqrt{M_x\pi}} \int \frac{ e^{- \tau_x^2 / M_x} \, H_{n_x+1}\left(\frac{\tau_x}{\sqrt{M_x}} \right) \, H_{n_x}\left( \frac{\tau_x}{\sqrt{M_x}} \right) \, e^{2\pi i q_x \tau_x} }{\sqrt{2^{n_x+1}\,(n_x+1)! 2^{n_x}\,n_x!}} d\tau_x \\
        = {} & {} \int \frac{ e^{- \tilde{x}^2} \, H_{n_x+1}\left(\tilde{x} \right) \, H_{n_x}\left( \tilde{x} \right) \, e^{2\pi i q_x \sqrt{M_x} \tilde{x}} }{\sqrt{2^{n_x+1} (n_x+1)! \; 2^{n_x}n_x! \pi}} d\tilde{x},
    \end{aligned}
    \end{equation}
where $M_x = \hbar / (M\omega_x)$ and $\tilde{x} = \tau_x / \sqrt{M_x}$. We can evaluate this integral for $n=n_x+1$, $m=n_x$, and $\beta_x = 2\pi \sqrt{M_x} q_x$ with the help of Eq.~\eqref{app_eq:int_e-x2_Hn_Hm_eibetax}, i.e.,
\begin{equation}
    \begin{aligned}
        \int \frac{ e^{- \tilde{x}^2} \, H_{n_x+1}\left(\tilde{x} \right) \, H_{n_x}\left( \tilde{x} \right) \, e^{ i \beta_x \tilde{x}} }{\sqrt{2^{n_x+1} (n_x+1)! \; 2^{n_x}n_x! \pi}} d\tilde{x} 
        = {} & {} \sqrt{\frac{1}{2(n_x+1)}} (i\beta_x) \; L_{n_x}^1 \left(\frac{\beta_x^2}{2}\right) \; e^{-\frac{\beta_x^2}{4}} \\
        = {} & {} i \pi q_x \frac{\sqrt{2 M_x}}{\sqrt{n_x+1}} \; L_{n_x}^1\left( 2 \pi^2 M_x q_x^2 \right) e^{- \pi^2 M_x q_x^2}.
    \end{aligned}
\end{equation}
We turn our attention to the second integral in eq.~\eqref{app_eq:QEP_AQHO_psi_nn+1_integrals_an_an+1_eiqtau_2ndterm}, i.e.,
\begin{equation}
    \label{app_eq:QEP_AQHO_integral_anny}
    \begin{aligned}
        \int a_{n_y}^*(\tau_y) {} & {} a_{n_y}(\tau_y) \, e^{2\pi i q_y \tau_y} d\tau_y 
        = \int \frac{ e^{- \tilde{y}^2} \, H_{n_y}\left(\tilde{y} \right) \, H_{n_y}\left( \tilde{y} \right) \, e^{i \beta_y \tilde{y}} }{\sqrt{2^{n_y} n_y! \; 2^{n_y}n_y!\pi}} d\tilde{y},
    \end{aligned}
\end{equation}
where, analogous to eq.~\eqref{app_eq:QEP_AQHO_integral_an+1n}, $M_y = \hbar/ M\omega_y$, $\tilde{y} = \tau_y / \sqrt{M_y}$, and $\beta_y = 2\pi \sqrt{M_y} q_y$. We can also use eq.~\eqref{app_eq:int_e-x2_Hn_Hm_eibetax} to evaluate the integral appearing in eq.~\eqref{app_eq:QEP_AQHO_integral_anny} and the result of such a procedure reads
    \begin{equation}
    \begin{aligned}
        \int a_{n_y}^*(\tau_y) {} & {} a_{n_y}(\tau_y) \, e^{2\pi i q_y \tau_y} d\tau_y 
        = L_{n_y}^0 \left(\frac{\beta_y^2}{2}\right) \; e^{-\frac{\beta_y^2}{4}} = L_{n_y}^0\left( 2 \pi^2 M_y q_y^2 \right) \; e^{- \pi^2 M_y q_y^2}.
    \end{aligned}
    \end{equation}
Putting these results together, we find for the inelastic wave associated with a transition $(n_x,n_y)\rightarrow (n_x+1,n_y)$
    \begin{equation}
        \label{app_eq:QEP_AQHO_psi_nn+1_final}
        \psi_{n_x, n_x+1}(\mathbf{q}) 
        = - \frac{\pi \gamma}{k_0 L} \; \frac{\sqrt{2 M_x}}{\sqrt{n_x+1}} \; q_x f_e(q) e^{- W_{0x}(q_x)} e^{- W_{0y}(q_y)} \;  L_{n_x}^1\big( 2 W_{0x}(q_x) \big) \; L_{n_y}^0\big( 2 W_{0y}(q_y) \big),
    \end{equation}
where we have made use of the DWF exponents $W_{0x}(q_x) = \pi^2 q_x^2 M_x$ and $W_{0y}(q_y) = \pi^2 q_y^2 M_y$ associated with zero-point motion as defined in appendix~\ref{app:transition_potentials_Forbes}. Equation~\eqref{app_eq:QEP_AQHO_psi_nn+1_final} is precisely eq.~\eqref{eq:QEP_AQHO_psi_nn+1_final}.

\section{Thermal average of the QEP cross section for an AQHO}
\label{app:QEP_thermal_average_AQHO}

We calculate the thermal average in Eq.~\eqref{eq:QEP_AQHO_ddscs_thermal_average_basic_expression} for the cross section of the process $n_x \rightarrow n_x +1$ in Eq.~\eqref{eq:QEP_AQHO_ddscs_nn+1_no_avg_general_init_state}, i.e.,
\begin{align}
    \label{app_eq:QEP_AQHO_ddscs_thermal_average_basic_expression}
    \frac{\mathrm{d}^2 \sigma (\mathbf{q}, \omega, T)}{\mathrm{d} \Omega_{\mathbf{q}} \mathrm{d} \omega}   
    = {} & {} \frac{1}{Z} \sum_{n_x,n_y} e^{-\beta E_{\mathbf{n}}} \frac{\mathrm{d} \sigma_{n_x, n_x+1} (\mathbf{q}, \omega)}{\mathrm{d} \Omega_{\mathbf{q}} \mathrm{d}\omega} \nonumber\\
    = {} & {} \frac{1}{Z} \sum_{n_x,n_y=0}^{\infty} e^{-\beta (n_x+\frac{1}{2}) \hbar\omega_{x}} e^{-\beta (n_y+\frac{1}{2}) \hbar\omega_{y}} \; \frac{\mathrm{d} \sigma_{n_x, n_x+1} (\mathbf{q}, \omega)}{\mathrm{d} \Omega_{\mathbf{q}} \mathrm{d}\omega}
\end{align}
The partition function is given by the expression
\begin{equation}
    \label{app_eq:QEP_AQHO_partition_function}
    Z 
    = \sum_{n_x,n_y=0}^{\infty} e^{-\beta (n_x+\frac{1}{2})\omega_{x}} e^{-\beta (n_y+\frac{1}{2})\omega_{y}}
    = \left[ z_x^{\frac{1}{2}} \sum_{n_x=0}^{\infty} z_x^{n_x} \right] \left[ z_y^{\frac{1}{2}} \sum_{n_y=0}^{\infty} z_y^{n_y} \right]
    = \frac{\sqrt{z}_x}{1-z_x} \frac{\sqrt{z_y}}{1-z_y},
\end{equation}
where $z_x = e^{-\beta \hbar\omega_{x}}$, $z_y = e^{-\beta \hbar\omega_{y}}$, and $\sum_{n=0}^{\infty} z^{n} = \frac{1}{1-z}$ is the geometric series for $z_y < 1$. In order to carry out the summation in Eq.~\eqref{app_eq:QEP_AQHO_ddscs_thermal_average_basic_expression}, we need to consider only those terms in Eq.~\eqref{eq:QEP_AQHO_ddscs_nn+1_no_avg_general_init_state} which explicitly depend on the occupation numbers $n_x$ and $n_y$, since all other factors will be just prefactors to the sum. Thus we need to evaluate
\begin{equation}
    \label{app_eq:statistical_sum_nn+1_simplified}
    \begin{aligned}
        \frac{1}{Z} \sum_{n_x,n_y=0}^{\infty} {} & {} e^{-\beta (n_x+\frac{1}{2}) \hbar\omega_{x}} e^{-\beta (n_y+\frac{1}{2}) \hbar\omega_{y}} \frac{\mathrm{d} \sigma_{n_x, n_x+1} (\mathbf{q}, \omega)}{\mathrm{d} \Omega_{\mathbf{q}}} \\
        {} & {} \propto
        \frac{\sqrt{z_x} \sqrt{z_y}}{Z} \sum_{n_x,n_y} z_x^{n_x} z_y^{n_y} \; \frac{1}{n_x+1} \;  \left\lbrace L_{n_x}^1\big( 2 W_{0x}(q_x) \big) \; L_{n_y}^0\big( 2 W_{0y}(q_y) \big) \right\rbrace^2.
    \end{aligned}
    \end{equation}
\end{widetext}

The sums over $n_x$ and $n_y$ can be carried out individually, since no factor depends on both occupation numbers. Exploiting Eq.~\eqref{app_eq:lagident} we find
\begin{equation}
\begin{aligned}
    \sqrt{z_x} \sum_{n_x=0}^{\infty} {} & {} \frac{1}{n_x+1} \; z_x^{n_x} \; \left\lbrace L_{n_x}^1 \big( 2 W_{0x}(q_x) \big) \right\rbrace^2 \\
    = {} & {} \frac{\sqrt{z_x}}{1-z_x} \; e^{-4 W_{0x}(q_x) \, \frac{z_x}{1-z_x}} \; \frac{I_1\big(4 W_{0x}(q_x) \frac{\sqrt{z_x}}{1-z_x} \big)}{\sqrt{4 W_{0x}(q_x)^2 z_x}},
\end{aligned}
\end{equation}
where $I_1$ is a modified Bessel function of the first kind. Similarly we compute the sum over $n_y$ in Eq.~\eqref{app_eq:statistical_sum_nn+1_simplified}, i.e.,
\begin{equation}
\begin{aligned}
    \sqrt{z_y} \sum_{n_y=0}^{\infty} {} & {} z_x^{n_y} \; \left\lbrace L_{n_y}^0\big( 2 W_{0y}(q_y) \big) \right\rbrace^2 \\
    = {} & {} \frac{\sqrt{z_y} }{1-z_y} \; e^{-4 W_{0y}(q_y) \frac{z_y}{1-z_y}} \; I_0\left(4 W_{0y}(q_y) \frac{\sqrt{z_y}}{1-z_y} \right).
\end{aligned}
\end{equation}

We put all of these results together, simplify using Eq.~\eqref{app_eq:QEP_AQHO_partition_function}, add all prefactors (which we neglected previously) and in this way obtain
\begin{eqnarray}
  \lefteqn{\frac{\mathrm{d}^2 \sigma (\mathbf{q}, \omega, T)}{\mathrm{d} \Omega_{\mathbf{q}} \mathrm{d} \omega}} \nonumber \\
  & = & \pi^2 \gamma^2 \; q_x^2 f_e^2(q) \; e^{- 2W_{0x}(q_x)} e^{- 2 W_{0y}(q_y)} \nonumber \\
  & \times & e^{-4 W_{0x}(q_x) \frac{z_x}{1-z_x}} \; e^{-4 W_{0y}(q_y) \frac{z_y}{1-z_y}} \nonumber \\
  & \times & 2 M_x \; \frac{I_1\left(4 W_{0x}(q_x) \frac{\sqrt{z_x}}{1-z_x} \right)}{2 W_{0x}(q_x) \sqrt{z_x}} \nonumber \\
  & \times & I_0\left(4 W_{0y}(q_y) \frac{\sqrt{z_y}}{1-z_y} \right) \delta(\omega-\omega_x).
\end{eqnarray}
We can further simplify this expression using Eq.~\eqref{app_eq:DWF_2D_W0},
\begin{equation*}
\begin{aligned}
    e^{- 2W_{0x}(q_x)} e^{-4 W_{0x}(q_x) \frac{z_x}{1-z_x}} 
    = {} & {} e^{- 2 W_{0x}(q_x) \left[1+ 2\frac{z_x}{1-z_x}\right]} \\
    = {} & {} e^{- 2W_{0x}(q_x) \coth(\frac{\beta\hbar\omega_x}{2})},
\end{aligned}
\end{equation*}
and then
\begin{equation}
    \begin{aligned}
        W_{0x}(q_x) \coth\left( \frac{\beta\hbar\omega_x}{2} \right) 
        = {} & {} \pi^2 M_x q_x^2 \coth\left(\frac{\beta\hbar\omega_x}{2}\right) \\
        = {} & {} 2 \pi^2 q_x^2 \left\langle u_x^2 \right\rangle \\
        = {} & {} W_{x}(q_x),
    \end{aligned}
\end{equation}
where we used Eq.~\eqref{app_eq:properties_QHO_MSD} and \eqref{app_eq:DWF_anisotropic_no_correlation}, and 
\begin{equation*}
    \frac{\sqrt{z_x}}{1-z_x} = \frac{e^{\frac{\beta\hbar\omega_x}{2}}}{1-e^{\beta\hbar\omega_x}} = \frac{1}{2} \sinh^{-1}\left(\frac{\beta\hbar\omega_x}{2}\right),
\end{equation*}
such that we end up with
\begin{eqnarray}\label{app_eq:QEP_AQHO_ddscs_thermal_average_final}
  \lefteqn{\frac{\mathrm{d}^2 \sigma (\mathbf{q}, \omega, T)}{\mathrm{d} \Omega_{\mathbf{q}} \mathrm{d} \omega}} \nonumber \\
  & = & \gamma^2 f_e^2(q) \; e^{- 2W_{x}(q_x)} e^{- 2 W_{y}(q_y)} e^{\frac{\beta\hbar\omega_x}{2}} \delta(\omega-\omega_x) \nonumber \\ 
  & \times & I_1\left(\frac{2 W_{0x}(q_x)}{\sinh\left(\frac{\beta\hbar\omega_x}{2}\right)} \right) \; I_0\left(\frac{2 W_{0y}(q_y)}{\sinh\left(\frac{\beta\hbar\omega_y}{2}\right)} \right).
\end{eqnarray}
Equation~\ref{app_eq:QEP_AQHO_ddscs_thermal_average_final} is the same as Eq.~\eqref{eq:QEP_AQHO_ddscs_thermal_average_final}.

\section{FRFPMS in WPOA - AQHO}
\label{app:FRFPMS_WPOA}

We derive here Eqs.~\eqref{eq:FRFPMS_AQHO_Iincoh}, \eqref{eq:FRFPMS_AQHO_Icoh} and \eqref{eq:FRFPMS_AQHO_Ivib}. We consider the 2D AQHO with incident plane wave in WPOA, so that the scattered wave is given by Eq.~\eqref{eq:WPOA_reciprocal_space}. We parametrize the displaced projected scattering potential associated with the AQHO as
\begin{equation}
    V_{\mathrm{proj}}(\mathbf{q}, \tau_x) = V_{\mathrm{proj}}(q) \; e^{2\pi i q_x \tau_x},
\end{equation}
where $\tau_x$ is the displacement from equilibrium. We consider the incoherent average prescribed by Eq.~\eqref{eq:FRFPMS_Iincoh}, i.e.,
\begin{eqnarray}
        \label{app_eq:FRFPMS_AQHO_Iincoh_start}
        I_{\mathrm{incoh}} (\mathbf{q})
        & = & \left\langle \left|\phi(\mathbf{q})\right|^2 \right\rangle_T \nonumber\\
        & = & \left\langle \left[\frac{\delta(\mathbf{q})}{L} + \frac{i\sigma}{L} V_{\mathrm{proj}}(\mathbf{q}, \tau_x) \right] \right. \nonumber\\
        & {} & \phantom{\left\langle \right.} \times \left.\left[ \frac{\delta(\mathbf{q})}{L^2} - \frac{i\sigma}{L} V_{\mathrm{proj}}^* (\mathbf{q}, \tau_x) \right] \right\rangle_T \nonumber\\
        & = & \frac{\delta(\mathbf{q})}{L^2} + \frac{\sigma^2}{L^2} \left\langle \left| V_{\mathrm{proj}}(\mathbf{q}, \tau_x) \right|^2 \right\rangle_T \nonumber \\
        & {} &  + \frac{i\sigma}{L^2} \delta(\mathbf{q}) \left\langle V_{\mathrm{proj}}(\mathbf{q}, \tau_x) \right\rangle_T \nonumber\\
        & {} &  - \frac{i\sigma}{L^2} \delta(\mathbf{q}) \left\langle V_{\mathrm{proj}}^{*} (\mathbf{q}, \tau_x) \right\rangle_T,
\end{eqnarray}
where we have assumed, that the displacement is the only variable, which is affected by the thermal averaging $\left\langle \ldots \right\rangle_T$. Equation~\eqref{app_eq:FRFPMS_AQHO_Iincoh_start} contains two types of averages over the potential, $\left\langle V(\mathbf{q}, \tau_x) \right\rangle_T$ and $\left\langle \left| V_{\mathrm{proj}}(\mathbf{q}, \tau_x) \right|^2 \right\rangle_T$. The former average can be evaluated according to
\begin{equation}
    \label{app_eq:FRFPMS_AQHO_DWF_derivation}
    \begin{aligned}
        \left\langle V(\mathbf{q}, \tau_x) \right\rangle_T
        = {} & {} V_{\mathrm{proj}}(q) \left\langle e^{2 \pi i q_x \tau_x} \right\rangle_T \\
        = {} & {} V_{\mathrm{proj}}(q) \; e^{-2\pi^2 q_x^2 \left\langle \tau_x^{2} \right\rangle_T} \\
        = {} & {} V_{\mathrm{proj}}(q) \; e^{-W_x(q_x)},
    \end{aligned}
\end{equation}
where we used Eq.~\eqref{app_eq:<expiqu>}.
Note that Eq.~\eqref{app_eq:FRFPMS_AQHO_DWF_derivation} expresses, that the averaged projected scattering potential due to a displacement along $x$ is just the projected scattering potential smeared by a DWF corresponding to the MSD in the $x$-direction, similarly to how we introduced the DWF in appendix~\eqref{app:DWF}. The average potential is real, i.e. $\left\langle V(\mathbf{q}, \tau_x) \right\rangle_T = \left\langle V^*(\mathbf{q}, \tau_x) \right\rangle_T$, since $V_{\mathrm{proj}}(q)$ is real (it is the Fourier transform of a real, even function $V(r)$), and the incoherent intensity simplifies, therefore, to
\begin{equation}
    \label{app_eq:FRFPMS_AQHO_Iincoh_final}
    I_{\mathrm{incoh}} (\mathbf{q}) = \frac{1}{L^2} \left[\delta(\mathbf{q}) + \sigma^2 \left\langle \left| V_{\mathrm{proj}}(\mathbf{q}, \tau_x) \right|^2 \right\rangle_T\right].
\end{equation}
Thus we are left to compute
\begin{equation}
    \begin{aligned}
        \left\langle \left| V_{\mathrm{proj}}(\mathbf{q}, \tau_x) \right|^2 \right\rangle_T
        = {} & {} \left\langle V_{\mathrm{proj}}(\mathbf{q}, \tau_x) \, V_{\mathrm{proj}}^* (\mathbf{q}, \tau_x) \right\rangle_T \\
        = {} & {} V_{\mathrm{proj}}^2(q) \left\langle e^{2\pi i q_x \tau_x} \, e^{-2\pi i q_x \tau_x} \right\rangle_T \\
        = {} & {} V_{\mathrm{proj}}^2(q),
    \end{aligned}
\end{equation}
since the displacement operator $\tau_x$ commutes with itself. Thus
\begin{equation}
    I_{\mathrm{incoh}} (\mathbf{q})
    = \frac{1}{L^2} \left[ \delta(\mathbf{q}) + \sigma^2 V_{\mathrm{proj}}^2(q) \right]
\end{equation}
and the coherent intensity prescribed by Eq.~\eqref{eq:FRFPMS_Icoh} becomes
\begin{eqnarray}
        \label{app_eq:FRFPMS_AQHO_Icoh_final}
        I_{\mathrm{coh}} (\mathbf{q}) 
        & = & \left\langle \left|\phi(\mathbf{q})\right|^2 \right\rangle_T \nonumber \\
        & = & \left\langle \left[\frac{\delta(\mathbf{q})}{L} + \frac{i\sigma}{L} V_{\mathrm{proj}}(\mathbf{q}, \tau_x) \right] \right\rangle_T \nonumber\\
        & {} & \times \left\langle \left[ \frac{\delta(\mathbf{q})}{L} - \frac{i\sigma}{L} V_{\mathrm{proj}}^* (\mathbf{q}, \tau_x) \right] \right\rangle_T \nonumber\\
        & = & \frac{\delta(\mathbf{q})}{L^2} + \frac{\sigma^2}{L^2} \left\langle V_{\mathrm{proj}}(\mathbf{q}, \tau_x) \right\rangle_T \left\langle V_{\mathrm{proj}}^* (\mathbf{q}, \tau_x) \right\rangle_T \nonumber \\
        & {} & + \frac{i\sigma^2}{L^2} \; \delta(\mathbf{q}) \left[ \left\langle V_{\mathrm{proj}}(\mathbf{q}, \tau_x) \right\rangle_T - \left\langle V_{\mathrm{proj}}^{*} (\mathbf{q}, \tau_x) \right\rangle_T \right] \nonumber\\
        & = & \frac{1}{L^2} \left[ \delta(\mathbf{q}) + \sigma^2 V_{\mathrm{proj}}^2(q) e^{-2W_x(q_x)} \right] \nonumber\\
        & = & \frac{1}{L^2} \left[ \delta(\mathbf{q}) + \frac{\gamma^2}{k_0^2} f_e^2(q) e^{-2W_x(q_x)} \right].
    \end{eqnarray}

Taking the difference of Eqs.~\eqref{app_eq:FRFPMS_AQHO_Iincoh_final} and \eqref{app_eq:FRFPMS_AQHO_Icoh_final} yields finally the vibrational intensity according to Eq.~\eqref{eq:FRFPMS_Ivib}, i.e.,
\begin{equation}
\begin{aligned}
    I_{\mathrm{vib}} 
    = {} & {} I_{\mathrm{incoh}} (\mathbf{q}) - I_{\mathrm{coh}} (\mathbf{q}) \\
    = {} & {} \sigma^2 V_{\mathrm{proj}}^2(q) \left[ 1 - e^{-2W_x(q_x)} \right] \\
    = {} & {} \sigma^2 V_{\mathrm{proj}}^2(q) e^{-2W_x(q_x)} \left[ e^{2W_x(q_x)} - 1 \right] \\
    = {} & {} \frac{\gamma^2}{k_0^2 L^2} f_e^2(q) e^{-2W_x(q_x)} \left[ e^{2W_x(q_x)} - 1 \right].
\end{aligned}
\end{equation}

\bibliography{references.bib}%

\end{document}